\documentclass{article}
\usepackage{amsmath}
\usepackage{graphicx}
\usepackage{color}
\usepackage{fullpage}
\usepackage{tikz}
\usetikzlibrary{arrows}
\newcommand{\Bi}{\mathop{\mathrm{Bi}}}
\newcommand{\NB}{\mathop{\mathrm{NB}}}

\newcommand{\ave}[1]{\left \langle #1 \right \rangle}
\newcommand{\pd}[2]{\frac{\partial #1}{\partial #2}}

\newcommand{\bx}{\boldsymbol{x}}
\newcommand{\bk}{\boldsymbol{k}}
\newcommand{\btheta}{\boldsymbol{\theta}}

\title{Edge-Based Compartmental Modeling for Infectious Disease Spread
  Part III: Disease and Population Structure}
\author{Joel C. Miller\footnote{Center for Communicable Disease Dynamics, Department of Epidemiology, Harvard School of Public Health}~\footnote{Fogarty International Center, NIH}\and Erik M. Volz\footnote{Department of Epidemiology, University of Michigan, Ann Arbor}}

\begin{document}
\maketitle

\begin{abstract}
We consider the edge-based compartmental models for infectious disease spread introduced in~\cite{miller:ebcm_overview}.  These models allow us to consider standard SIR diseases spreading in random populations.  In this paper we show how to handle deviations of the disease or population from the simplistic assumptions of~\cite{miller:ebcm_overview}.  We allow the population to have structure due to effects such as demographic detail or multiple types of risk behavior the disease to have more complicated natural history.  We introduce these modifications in the static network context, though it is straightforward to incorporate them into dynamic networks.  We also consider serosorting, which requires using the dynamic network models. The basic methods we use to derive these generalizations are widely applicable, and so it is straightforward to introduce many other generalizations not considered here.
\end{abstract}

\section{Introduction}
In~\cite{miller:ebcm_overview}, we introduced the edge-based compartmental modeling approach for infectious disease spread.  We showed how to create models that account for social heterogeneity and contact duration simultaneously.  However, there are many other important effects that govern the spread of disease through a population which were not considered.   For example it is well-established that for diseases such as influenza, heterogeneities in infectiousness/susceptibility and biases in mixing among age groups plays an important role~\cite{wallinga:contact_survey}.  Further, it is widely believed that for some diseases there are different stages of infection that have population scale impacts such as a highly infectious early phase followed by a chronic less infectious phase as in HIV~\cite{pilcher:acute_hiv} or a latent uninfectious early phase followed by an infection active phase as in Tuberculosis.  If there is structure within a population or within the progression of disease is important for our models to incorporate the details.

Many of these effects have been studied under mass action mixing assumptions.  In this paper we show that a range of assumptions can be reliably and simply captured by edge-based compartmental models.  We begin by rederiving the simplest such model for Configuration Model networks.  
We then look at a number of different population or disease structures, summarized in Table~\ref{tab:models}, and show how to develop an edge-based compartmental model incorporating these effects.  The effects in the list are chosen to demonstrate the flexibility of the approach.  It is by no means exhaustive, and it is not difficult to combine different effects.  These model populations are mostly static, and we describe them in the Configuration Model case, but it is straightforward to adapt these effects to other static or dynamic networks.  Because of its epidemiological importance, we end the paper by showing a model incorporating serosorting where the population structure changes in response to the disease.  This is inherently a dynamic network process, so we must use a dynamic network model to study it.

\begin{table}
\newcommand\T{\rule{0pt}{2.4ex}}
\newcommand\B{\rule[-1.1ex]{0pt}{0pt}}
\begin{center}
\begin{tabular}{|c|c|c|}
\hline
\T{}\textbf{Model}\B{}   &   \textbf{Brief Description}  &  \textbf{Section} \\\hline\hline%
\parbox{0.25\textwidth}{\footnotesize Directed Networks}&\parbox{0.6\textwidth}{\footnotesize \T{}Model for a disease in which some contacts are not symmetric in terms of disease risk.\B{}}&\ref{sec:directed}\\\hline%
\parbox{0.25\textwidth}{\footnotesize Heterogeneous Individuals}&\parbox{0.6\textwidth}{\footnotesize \T{}Model for populations with heterogeneities in infectiousness and/or susceptibility that do not correlate with population structure \B{}}&\ref{sec:heterog}\\\hline%
\parbox{0.25\textwidth}{\footnotesize Assortative mixing by type}&\parbox{0.6\textwidth}{\footnotesize \T{}Model for populations with demographic groups that have heterogeneities in infectiousness and/or susceptibility where contacts are affected by an individual's group.\B{}}&\ref{sec:demog}\\\hline %
\parbox{0.25\textwidth}{\footnotesize Multiple transmission modes}&\parbox{0.6\textwidth}{\footnotesize \T{}Model for a disease that can be transmitted by more than one type of behavior and the network structure induced by each behavior is different.\B{}}&\ref{sec:multimode}\\\hline %
\parbox{0.25\textwidth}{\footnotesize Multiple infectious stages}&\parbox{0.6\textwidth}{\footnotesize \T{}Model for a disease which has several infectious stages of possibly varying duration or infectiousness.\B{}}&\ref{sec:multistage}\\\hline %
%
%
%
%
\parbox{0.25\textwidth}{\footnotesize Serosorting }&\parbox{0.6\textwidth}{\footnotesize \T{}Model for dynamic network where edges break or are created at rates dependent on the status of partners.\B{}}&\ref{sec:fd_serosort}, \ref{sec:vd_serosort} \\\hline%
\end{tabular}
\end{center}
\caption{The edge-based compartmental models considered here.  All of these except serosorting are presented using the (static) Configuration Model network structure.  Serosorting is presented in two different dynamic network contexts.}
\label{tab:models}
\end{table}

\section{The basic model}
The static models we present here are all generalizations of the basic Configuration Model (CM) static network epidemic model.  To set the stage, we first define a CM network~\cite{MolloyReed,newman:structurereview,newman:arb_degree,vanderhofstad:randomgraphs}.  Because we will use the same underlying approach, we also briefly describe the method of~\cite{miller:ebcm_overview}.

In a CM network, each individual is represented by a \emph{node} which is connected to other nodes by \emph{edges} which can transmit disease.  To construct the network, each node is assigned a number of edges (it's \emph{degree}) $k$ with probability $P(k)$.  The edges connect randomly to other neighbors using proportional mixing, so that the probability of selecting a neighbor of degree $k$ is $k P(k)/\ave{K}$ where $\ave{K}$ denotes the average of $k$.  It is convenient to define
\[
\psi(x) = \sum_k P(k) x^k
\]
Note that $\psi'(x) = \sum_k k P(k) x^{k-1}$ and $\psi'(1) = \ave{K}$.

\begin{figure}
\parbox{\textwidth}{\parbox{0.5\textwidth}{\scalebox{0.95}{
\input{edgeflux_structure_paper.txt}
}} \hfill
\parbox{0.5\textwidth}{\scalebox{0.95}{
\begin{picture}(0,0)%
\includegraphics{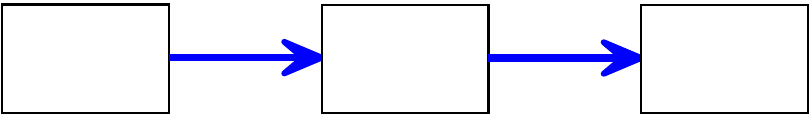}%
\end{picture}%
\setlength{\unitlength}{3947sp}%
\begingroup\makeatletter\ifx\SetFigFontNFSS\undefined%
\gdef\SetFigFontNFSS#1#2#3#4#5{%
  \reset@font\fontsize{#1}{#2pt}%
  \fontfamily{#3}\fontseries{#4}\fontshape{#5}%
  \selectfont}%
\fi\endgroup%
\begin{picture}(3883,550)(1651,-1886)
\put(2038,-1700){\makebox(0,0)[b]{\smash{{\SetFigFontNFSS{9}{9.6}{\familydefault}{\mddefault}{\updefault}{\color[rgb]{0,0,0}$S=\psi(\theta)$}%
}}}}
\put(3589,-1700){\makebox(0,0)[b]{\smash{{\SetFigFontNFSS{9}{9.6}{\familydefault}{\mddefault}{\updefault}{\color[rgb]{0,0,0}$I$}%
}}}}
\put(5140,-1700){\makebox(0,0)[b]{\smash{{\SetFigFontNFSS{9}{9.6}{\familydefault}{\mddefault}{\updefault}{\color[rgb]{0,0,0}$R$}%
}}}}
\put(4382,-1509){\makebox(0,0)[b]{\smash{{\SetFigFontNFSS{9}{9.6}{\familydefault}{\mddefault}{\updefault}{\color[rgb]{0,0,0}$\gamma I$}%
}}}}
\end{picture}%
}}} 
\caption{\textbf{Flow diagram for Configuration Model networks.}  
To find $\theta$, we must find $\phi_I$.  We do this by calculating $\phi_S$ explicitly which is $\psi'(\theta)/\psi'(1)$.  Because the flux into $\phi_R$ and $1-\theta$ are proportional, we can find $\phi_R$ in terms of $\theta$.  We then use $\phi_I = \theta-\phi_S-\phi_R$, and we are able to find a differential equation for $\theta$ in terms of $\theta$.}
\label{fig:CM_flow}
\end{figure}

When we considered the spread of epidemics in CM networks in our original paper~\cite{miller:ebcm_overview}, we used the flow diagram of figure~\ref{fig:CM_flow}.  Here $S(t)$ is the proportion of the population still susceptible, $I(t)$ is the proportion infected, and $R(t)$ the proportion recovered. Assuming we know $S(t)$, the probability a random individual is susceptible, we are able to determine $I$ and $R$ by $I=1-S-R$, \ $\dot{R} = \gamma I$.  

To calculate the probability a random individual is susceptible, we choose a random \emph{test node} $u$ uniformly from the population.  We alter $u$ so that if infected, $u$ does not transmit to its neighbors.  This helps us assume the status of its neighbors are independent, but does not affect the probability $u$ is susceptible.  We use $\theta$ to be the probability that a random neighbor $v$  of $u$ has not transmitted infection to $u$, and we break $\theta$ into three parts: $\phi_S$, $\phi_I$, and $\phi_R$, which are the probability $v$ is still susceptible, the probability $v$ is infected but has not transmitted to $u$, and the probability $v$ is recovered and did not transmit to $u$.  Because the infection rate is $\beta$ and the recovery rate is $\gamma$ it is relatively straightforward to see that the fluxes from $\phi_I$ to $1-\theta$ and $\phi_R$ are $\beta\phi_I$ and $\gamma \phi_I$ respectively.  The calculation of the flux from $\phi_S$ to $\phi_I$ is less obvious.  We have two options.  We can calculate the flux directly or we can calculate $\phi_S$ explicitly as was done in~\cite{miller:ebcm_overview}.  The first option is more general, but the second is simpler.

The flow diagram of figure~\ref{fig:CM_flow} gives
\begin{equation}
\dot{\theta} = -\beta \phi_I
\label{eqn:thetadot}
\end{equation}
The remainder of our derivation focuses on finding $\phi_I$.
Because $\theta=\phi_S+\phi_I+\phi_R$, we have $\phi_I = \theta - \phi_S - \phi_R$.  Thus we simply need to calculate $\phi_S$ and $\phi_R$ in terms of $\theta$ to find $\phi_I$ in terms of $\theta$.   To calculate $\phi_S$, we use the fact that the probability a neighbor $v$ has degree $k$ is $k P(k)/\ave{K}$.  Since $u$ is prevented from infecting $v$, the probability $v$ is susceptible given its $k$ is $\theta^{k-1}$.  Thus $\phi_S$ is a weighted average of this, $\phi_S = \sum_k k P(k) \theta^{k-1}/\ave{K} = \psi'(\theta)/\psi'(1)$.

To calculate $\phi_R$, we look at figure~\ref{fig:CM_flow}.  The fluxes into $\phi_R$ and into $1-\theta$ are proportional to one another, with the proportionality coefficient $\gamma/\beta$.  Since they both begin at $0$, this means $\phi_R = \gamma(1-\theta)/\beta$.  Consequently
\[
\phi_I = \theta - \frac{\psi'(\theta)}{\psi'(1)} - \frac{\gamma}{\beta} (1-\theta)
\]
This leads to 
\[
\dot{\theta} = -\beta\theta + \beta \frac{\psi'(\theta)}{\psi'(1)} + \gamma (1-\theta)
\]
Thus we have the system of equations
\begin{align}
\dot{\theta} &= -\beta\theta + \beta \frac{\psi'(\theta)}{\psi'(1)} + \gamma (1-\theta) \label{eqn:thetadot_final}\\
\dot{R} &= \gamma I  \qquad \qquad S = \psi(\theta) \qquad \qquad  I  = 1-S-R
\label{eqn:Rdot_final}
\end{align}
which has just two ordinary differential equations (ODEs).  In fact~\eqref{eqn:thetadot_final} does not depend on~\eqref{eqn:Rdot_final}, so the system is governed by the single ODE~\eqref{eqn:thetadot_final}.  

\subsection{Generalizing the model}
In the following, we generalize the model for many static network situations.  The basic approach is to consider a random test node which is prevented from causing infection.  Then consider the edges which could transmit infection to it.  We determine the probability the edges have not transmitted, which may depend on the test node, the neighbor, or details of the contact.  Our approach to determining this probability is the same as above.  Once we know the probability any given edge has not transmitted to the test node, we can calculate the probability that the test node is susceptible, from which we can calculate the proportion of the population that is susceptible, infected, or recovered.

\section{Epidemics on generalized static networks}
\subsection{Directed Networks}
\label{sec:directed}
There are a number of realistic scenarios where infection can transmit in only one direction.  Examples include blood transfusions, a restaurant cook infecting a patron, and even a patient infecting a doctor where they come into contact only because of the patient's infection.  The probability and final size of epidemics for this scenario have been studied previously~\cite{meyers:directed}, but not the dynamics.

\begin{figure}
\scalebox{0.95}{
\begin{picture}(0,0)%
\includegraphics{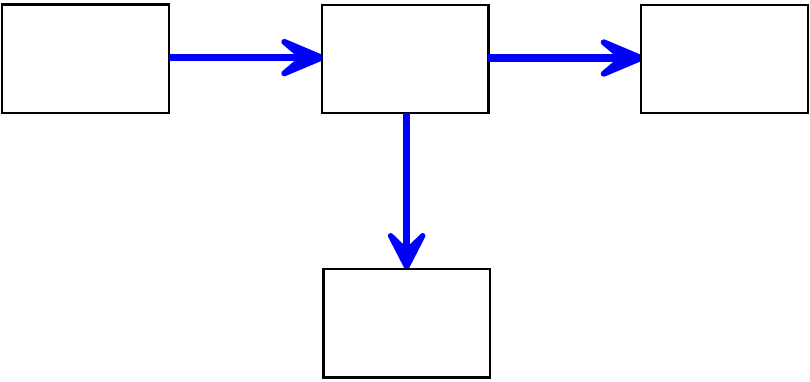}%
\end{picture}%
\setlength{\unitlength}{3947sp}%
\begingroup\makeatletter\ifx\SetFigFontNFSS\undefined%
\gdef\SetFigFontNFSS#1#2#3#4#5{%
  \reset@font\fontsize{#1}{#2pt}%
  \fontfamily{#3}\fontseries{#4}\fontshape{#5}%
  \selectfont}%
\fi\endgroup%
\begin{picture}(3883,1833)(2026,-3544)
\put(2149,-2047){\makebox(0,0)[b]{\smash{{\SetFigFontNFSS{9}{9.6}{\familydefault}{\mddefault}{\updefault}{\color[rgb]{0,0,0}\parbox{0.02\textwidth}{$\phi_{S,d}=\frac{\psi_y(\theta_d,1,\theta_n)}{\psi_y(1,1,1)}$}}%
}}}}
\put(3974,-2047){\makebox(0,0)[b]{\smash{{\SetFigFontNFSS{9}{9.6}{\familydefault}{\mddefault}{\updefault}{\color[rgb]{0,0,0}$\phi_{I,d}$}%
}}}}
\put(3974,-3310){\makebox(0,0)[b]{\smash{{\SetFigFontNFSS{9}{9.6}{\familydefault}{\mddefault}{\updefault}{\color[rgb]{0,0,0}$1-\theta_d$}%
}}}}
\put(5511,-2047){\makebox(0,0)[b]{\smash{{\SetFigFontNFSS{9}{9.6}{\familydefault}{\mddefault}{\updefault}{\color[rgb]{0,0,0}$\phi_{R,d}$}%
}}}}
\put(4732,-2224){\makebox(0,0)[b]{\smash{{\SetFigFontNFSS{9}{9.6}{\familydefault}{\mddefault}{\updefault}{\color[rgb]{0,0,0}$\gamma\phi_{I,d}$}%
}}}}
\put(4291,-2774){\makebox(0,0)[b]{\smash{{\SetFigFontNFSS{9}{9.6}{\familydefault}{\mddefault}{\updefault}{\color[rgb]{0,0,0}$\beta_d\phi_{I,d}$}%
}}}}
\end{picture}%
}\hfill\scalebox{0.95}{
\begin{picture}(0,0)%
\includegraphics{edgeflux.pdf}%
\end{picture}%
\setlength{\unitlength}{3947sp}%
\begingroup\makeatletter\ifx\SetFigFontNFSS\undefined%
\gdef\SetFigFontNFSS#1#2#3#4#5{%
  \reset@font\fontsize{#1}{#2pt}%
  \fontfamily{#3}\fontseries{#4}\fontshape{#5}%
  \selectfont}%
\fi\endgroup%
\begin{picture}(3883,1833)(2026,-3544)
\put(2149,-2047){\makebox(0,0)[b]{\smash{{\SetFigFontNFSS{9}{9.6}{\familydefault}{\mddefault}{\updefault}{\color[rgb]{0,0,0}\parbox{0.02\textwidth}{$\phi_{S,n}=\frac{\psi_z(\theta_d,1,\theta_n)}{\psi_z(1,1,1)}$}}%
}}}}
\put(3974,-2047){\makebox(0,0)[b]{\smash{{\SetFigFontNFSS{9}{9.6}{\familydefault}{\mddefault}{\updefault}{\color[rgb]{0,0,0}$\phi_{I,n}$}%
}}}}
\put(3974,-3310){\makebox(0,0)[b]{\smash{{\SetFigFontNFSS{9}{9.6}{\familydefault}{\mddefault}{\updefault}{\color[rgb]{0,0,0}$1-\theta_n$}%
}}}}
\put(5511,-2047){\makebox(0,0)[b]{\smash{{\SetFigFontNFSS{9}{9.6}{\familydefault}{\mddefault}{\updefault}{\color[rgb]{0,0,0}$\phi_{R,n}$}%
}}}}
\put(4732,-2224){\makebox(0,0)[b]{\smash{{\SetFigFontNFSS{9}{9.6}{\familydefault}{\mddefault}{\updefault}{\color[rgb]{0,0,0}$\gamma\phi_{I,n}$}%
}}}}
\put(4291,-2774){\makebox(0,0)[b]{\smash{{\SetFigFontNFSS{9}{9.6}{\familydefault}{\mddefault}{\updefault}{\color[rgb]{0,0,0}$\beta_n\phi_{I,n}$}%
}}}}
\end{picture}%
}\\[10pt]
\begin{center}
\parbox{0.5\textwidth}{\scalebox{0.95}{
\begin{picture}(0,0)%
\includegraphics{standardflux.pdf}%
\end{picture}%
\setlength{\unitlength}{3947sp}%
\begingroup\makeatletter\ifx\SetFigFontNFSS\undefined%
\gdef\SetFigFontNFSS#1#2#3#4#5{%
  \reset@font\fontsize{#1}{#2pt}%
  \fontfamily{#3}\fontseries{#4}\fontshape{#5}%
  \selectfont}%
\fi\endgroup%
\begin{picture}(3883,550)(1651,-1886)
\put(1768,-1700){\makebox(0,0)[b]{\smash{{\SetFigFontNFSS{9}{9.6}{\familydefault}{\mddefault}{\updefault}{\color[rgb]{0,0,0}\parbox{0.02\textwidth}{$S=\psi(\theta_d,1,\theta_n)$}}%
}}}}
\put(3589,-1700){\makebox(0,0)[b]{\smash{{\SetFigFontNFSS{9}{9.6}{\familydefault}{\mddefault}{\updefault}{\color[rgb]{0,0,0}$I$}%
}}}}
\put(5140,-1700){\makebox(0,0)[b]{\smash{{\SetFigFontNFSS{9}{9.6}{\familydefault}{\mddefault}{\updefault}{\color[rgb]{0,0,0}$R$}%
}}}}
\put(4382,-1509){\makebox(0,0)[b]{\smash{{\SetFigFontNFSS{9}{9.6}{\familydefault}{\mddefault}{\updefault}{\color[rgb]{0,0,0}$\gamma I$}%
}}}}
\end{picture}%
}}
\end{center}
\caption{\textbf{Directed CM model.} Flow diagram for a network with directed and nondirected edges.  We consider the two edge types separately.  The evolution of edges is similar to figure~\ref{fig:CM_flow}.  We can assign different infection rates for each edge type.}
\label{fig:direct}
\end{figure}

We can investigate the dynamics in almost the same manner as before.   Assume that the network consists of both directed and nondirected edges.  The disease can transmit along a directed edge only following the edge direction, while nondirected edges may be followed in either direction.  Let $\beta_d$ and $\beta_n$ denote the rate of transmission along directed and nondirected edges respectively.  Recovery occurs at rate $\gamma$ regardless of how infection is transmitted.

We refer to edges pointing to a node of interest as \emph{in-directed}, and those pointing away as \emph{out-directed} edges.  The probability of having $k_i$ in-, $k_o$ out-, and $k_n$ nondirected edges is given by $P(k_i,k_o,k_n)$.  We define
\[
\psi(x,y,z) = \sum_{k_i,k_o,k_n} P(k_i,k_o,k_n)x^{k_i}y^{k_o}z^{k_n}
\]
We again consider a random test node $u$ which is prevented from causing infection.  We define $\theta_d(t)$ and $\theta_n(t)$ to be the probability an in-directed edge or nondirected edge to $u$ respectively has not transmitted infection to $u$.  The probability that $u$ is still susceptible is $\psi(\theta_d,1,\theta_n)$.   We use the variables $\phi_{S,d}$, $\phi_{I,d}$, and $\phi_{R,d}$ to be the equivalent of $\phi_S$, $\phi_I$, and $\phi_R$ seen before for in-directed edges.  Following the same approach as before we arrive at the flow diagrams in figure~\ref{fig:direct}.  

Consider a neighbor $v$ with a directed edge to $u$.  
Because of how $v$ is chosen, the probability it has $k_i$ in-, $k_o$ out-, and $k_n$ nondirected edges is $k_oP(k_i,k_o,k_n)/\ave{K_o}$.   The probability that $v$ is still susceptible is $\phi_{S,d}=\sum_{k_i,k_o,k_n} k_o P(k_i,k_o,k_n)\theta_d^{k_i}\theta_n^{k_n}/\ave{K_o} = \pd{}{y}\psi(\theta_d,1,\theta_n)/\pd{}{y}\psi(1,1,1)$.  The probability that $v$ has recovered without infecting $u$ is $\phi_{R,d}=\gamma(1-\theta_d)/\beta_d$.  Because $\phi_{I,d} = \theta_d-\phi_{S,d}-\phi_{R,d}$ we  have $\phi_{I,d}$ in terms of $\theta_d$ and $\theta_n$.  This gives us $\dot{\theta}_d$ in terms of $\theta_d$ and $\theta_n$.  A similar expression holds for $\dot{\theta}_n$.  We have
\begin{align}
\dot{\theta}_d &= - \beta_d \theta_d + \beta_d \frac{\pd{}{y}\psi(\theta_d,1,\theta_n)}{\pd{}{y}\psi(1,1,1)}+ \gamma(1-\theta_d)\label{eqn:thetad}\\
\dot{\theta}_n &= - \beta_n \theta_n + \beta_n \frac{\pd{}{z}\psi(\theta_d,1,\theta_n)}{\pd{}{z}\psi(1,1,1)}+ \gamma (1-\theta_n)
\label{eqn:thetau}
\end{align}
To this we add
\begin{equation}
  \dot{R} = \gamma I \qquad\qquad
  S = \psi(\theta_d,1,\theta_n)\qquad\qquad
I = 1-S-R
\end{equation}
to give us the proportion of susceptible, infected, and recovered individuals.

If the system only has directed edges, then we can drop $\theta_n$ from the analysis and $\psi(x,y,z)$ reduces to $\psi(x,y)$.  Such a model could be used to study the impact of superspreaders where the probability of receiving infection from an infected node is similar for all nodes (in-degrees are similar), but some nodes have many more neighbors to infect than others (high variance in out-degree).

\subsubsection{Example}
\label{sec:directed_ex}
\begin{figure}
\begin{center}
\includegraphics[width=0.48\textwidth]{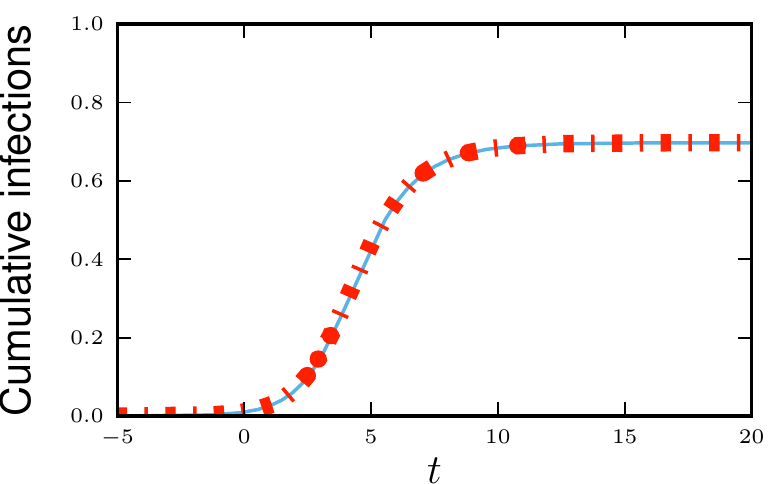}\hfill
\includegraphics[width=0.48\textwidth]{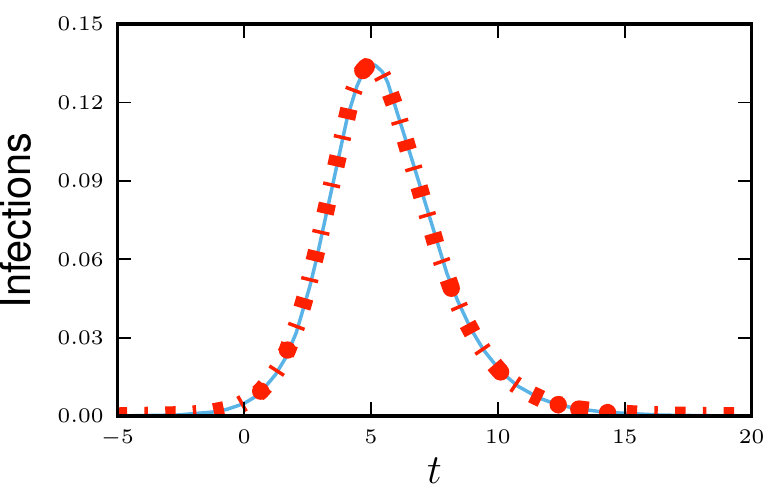}
\end{center}
\caption{\textbf{Directed CM example.} Results for the directed networks described in section~\ref{sec:directed_ex} using $\hat{k}=4$, \ $\beta_d=0.2$, and $\beta_n=0.4$.  We choose $t=0$ to correspond to $1\%$ cumulative incidence.  Theory (dashed) correctly predicts dynamics of simulations in a  population of $5\times 10^5$ individuals (solid).}
\label{fig:direx}
\end{figure}

Consider a population for which the average in-degree, out-degree, and nondirected degrees are each $\hat{k}$ as follows:  Each node has in-degree $\hat{k}$.  The out-degree $k_o$ is uniformly chosen from $0$ up to $2\hat{k}$ inclusive, and the nondirected degree is $2\hat{k}-k_o$.  For this population, 
\[
\psi(x,y,z) = x^{\hat{k}} \frac{z^{2\hat{k}+1}-y^{2\hat{k}+1}}{(2\hat{k}+1)(z-y)} \, .
\]  
Figure~\ref{fig:direx} shows results for $\hat{k}=4$, \ $\beta_d=0.2$, and $\beta_n=0.4$.

\subsection{Heterogeneous infectiousness and susceptibility}
\label{sec:heterog}
\begin{figure}
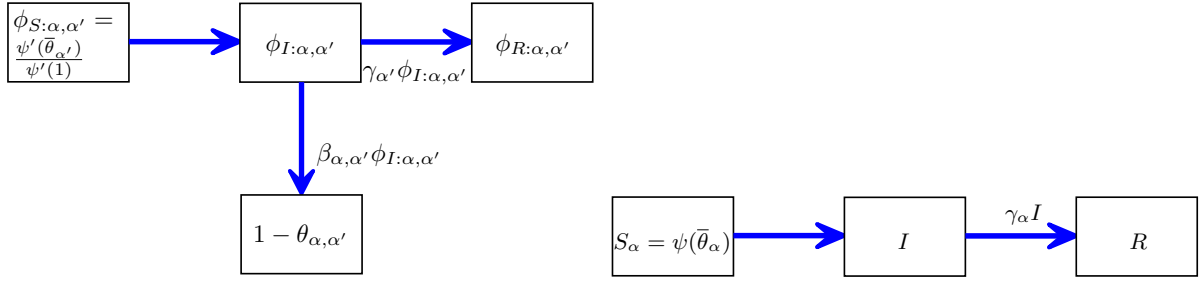

\begin{center}
\scalebox{0.95}{
\begin{picture}(0,0)%
\includegraphics{edgeflux.pdf}%
\end{picture}%
\setlength{\unitlength}{3947sp}%
\begingroup\makeatletter\ifx\SetFigFontNFSS\undefined%
\gdef\SetFigFontNFSS#1#2#3#4#5{%
  \reset@font\fontsize{#1}{#2pt}%
  \fontfamily{#3}\fontseries{#4}\fontshape{#5}%
  \selectfont}%
\fi\endgroup%
\begin{picture}(3883,1833)(2026,-3544)
\put(2149,-2047){\makebox(0,0)[b]{\smash{{\SetFigFontNFSS{10}{9.6}{\familydefault}{\mddefault}{\updefault}{\color[rgb]{0,0,0}\parbox{0.02\textwidth}{$\phi_{S:\alpha,\alpha'}=\frac{\psi'(\overline{\theta}_{\alpha'})}{\psi'(1)}$}}%
}}}}
\put(3974,-2047){\makebox(0,0)[b]{\smash{{\SetFigFontNFSS{10}{9.6}{\familydefault}{\mddefault}{\updefault}{\color[rgb]{0,0,0}$\phi_{I:\alpha,\alpha'}$}%
}}}}
\put(3974,-3310){\makebox(0,0)[b]{\smash{{\SetFigFontNFSS{10}{9.6}{\familydefault}{\mddefault}{\updefault}{\color[rgb]{0,0,0}$1-\theta_{\alpha,\alpha'}$}%
}}}}
\put(5511,-2047){\makebox(0,0)[b]{\smash{{\SetFigFontNFSS{10}{9.6}{\familydefault}{\mddefault}{\updefault}{\color[rgb]{0,0,0}$\phi_{R:\alpha,\alpha'}$}%
}}}}
\put(4732,-2224){\makebox(0,0)[b]{\smash{{\SetFigFontNFSS{10}{9.6}{\familydefault}{\mddefault}{\updefault}{\color[rgb]{0,0,0}$\gamma_{\alpha'}\phi_{I:\alpha,\alpha'}$}%
}}}}
\put(4491,-2774){\makebox(0,0)[b]{\smash{{\SetFigFontNFSS{10}{9.6}{\familydefault}{\mddefault}{\updefault}{\color[rgb]{0,0,0}$\beta_{\alpha,\alpha'}\phi_{I:\alpha,\alpha'}$}%
}}}}
\end{picture}%
} \scalebox{0.95}{
\begin{picture}(0,0)%
\includegraphics{standardflux.pdf}%
\end{picture}%
\setlength{\unitlength}{3947sp}%
\begingroup\makeatletter\ifx\SetFigFontNFSS\undefined%
\gdef\SetFigFontNFSS#1#2#3#4#5{%
  \reset@font\fontsize{#1}{#2pt}%
  \fontfamily{#3}\fontseries{#4}\fontshape{#5}%
  \selectfont}%
\fi\endgroup%
\begin{picture}(3883,550)(1651,-1886)
\put(2058,-1700){\makebox(0,0)[b]{\smash{{\SetFigFontNFSS{9}{9.6}{\familydefault}{\mddefault}{\updefault}{\color[rgb]{0,0,0}$S_\alpha = \psi(\overline{\theta}_\alpha)$}%
}}}}
\put(3589,-1700){\makebox(0,0)[b]{\smash{{\SetFigFontNFSS{9}{9.6}{\familydefault}{\mddefault}{\updefault}{\color[rgb]{0,0,0}$I$}%
}}}}
\put(5140,-1700){\makebox(0,0)[b]{\smash{{\SetFigFontNFSS{9}{9.6}{\familydefault}{\mddefault}{\updefault}{\color[rgb]{0,0,0}$R$}%
}}}}
\put(4382,-1509){\makebox(0,0)[b]{\smash{{\SetFigFontNFSS{9}{9.6}{\familydefault}{\mddefault}{\updefault}{\color[rgb]{0,0,0}$\gamma_\alpha I$}%
}}}}
\end{picture}%
}
\end{center}
\caption{\textbf{Heterogeneous infectiousness/susceptibility model.}  We separate nodes by type $\alpha$, but assume that $\alpha$ has no effect on connectivity.  Both infectiousness and susceptibility may depend on $\alpha$.  We must consider edges between each pair of types $\alpha$ and $\alpha'$ separately.  The evolution of edges is similar to before.}
\label{fig:het}
\end{figure}
 
Assume now that there is a parameter $\alpha$ which measures a node's ability to become infected and cause infection, but does not influence the contact structure of the population.  We refer to the value of $\alpha$ for a node as its \emph{type}, and the probability a node has a given type $\alpha$ is $Q(\alpha)$.  The recovery rate $\gamma_\alpha$ of a type-$\alpha$ node and the transmission rate $\beta_{\alpha,\alpha'}$ from a type-$\alpha'$ node to a type-$\alpha$ node are type dependent.  The final size for a special case of this model where a node's infectiousness and susceptibility are uncorrelated is given in~\cite{miller:heterogeneity}.

Consider a random test node $u$ of type $\alpha$.  Let $\theta_{\alpha,\alpha'}$ denote the probability that an edge from a  type-$\alpha'$ neighbor $v$ to $u$ has not transmitted infection from $v$ to $u$, and similarly $\phi_{S:\alpha,\alpha'}$ the probability $v$ is still susceptible, $\phi_{I:\alpha,\alpha'}$ the probability $v$ is infected but the edge has not transmitted, and $\phi_{R:\alpha,\alpha'}$ the probability that $v$ has recovered without transmitting.   We define $\overline{\theta}_{\alpha} = \sum_{\alpha'} \theta_{\alpha,\alpha'} Q(\alpha')$ as the probability that a random edge to $u$ has not transmitted infection to $u$.  We use the original definition of $\psi(x)=\sum_k P(k)x^k$.  

We find that $v$ is susceptible with probability $\phi_{S:\alpha,\alpha'} = \sum_k kP(k)  \overline{\theta}_{\alpha'}^{k-1} /\ave{K} = \psi'(\overline{\theta}_{\alpha'})/\psi'(1)$.  We also find that $\phi_{R:\alpha,\alpha'} = \gamma_{\alpha'}(1-\theta_{\alpha,\alpha'})/\beta_{\alpha,\alpha'}$.
Then the flow diagram in figure~\ref{fig:het} shows that
\begin{equation}
\dot{\theta}_{\alpha,\alpha'} = -\beta_{\alpha,\alpha'}\theta_{\alpha,\alpha'} + \beta_{\alpha,\alpha'} \frac{\psi'(\overline{\theta}_{\alpha'})}{\psi'(1)} +  \gamma_{\alpha'}(1-\theta_{\alpha,\alpha'}) 
\label{eqn:theta12}
\end{equation}
The probabilities a type-$\alpha$ node is still susceptible, infected, or recovered satisfy
\begin{equation}
\dot{R}_\alpha = \gamma_\alpha I_\alpha \qquad\qquad 
S_\alpha = \psi(\overline{\theta}_\alpha) \qquad\qquad
I_\alpha = 1-S_\alpha -R_\alpha 
\end{equation}
The total population in each state is given by
\begin{equation}
S = \sum_\alpha S_\alpha Q(\alpha) \qquad\qquad I = \sum_\alpha I_\alpha Q(\alpha) \qquad\qquad
R = \sum_\alpha R_\alpha Q(\alpha) 
\end{equation}


\subsubsection{Example}
\begin{figure}
\begin{center}
\includegraphics[width=0.48\textwidth]{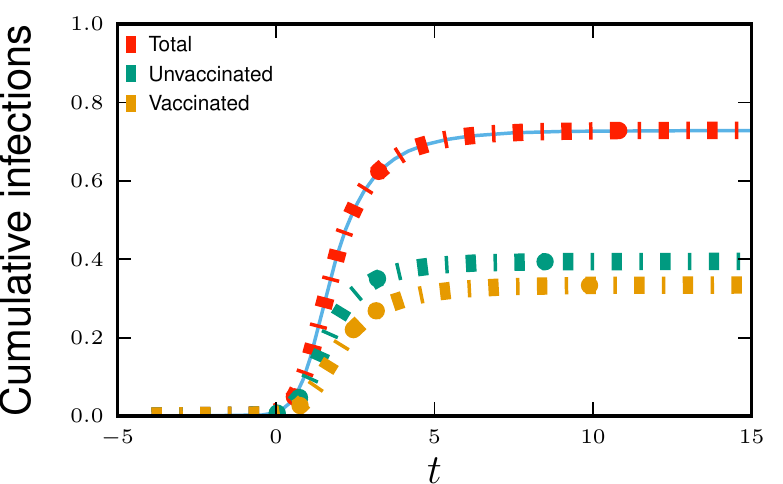} \hfill
\includegraphics[width=0.48\textwidth]{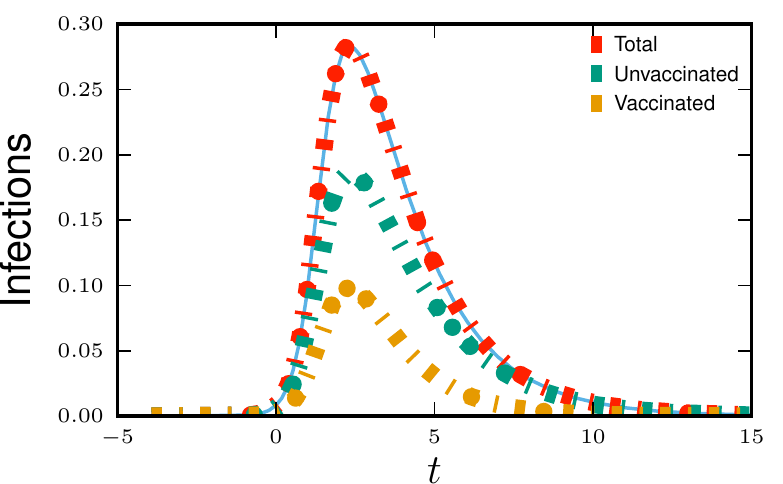}
\end{center}
\caption{\textbf{Heterogeneous infectiousness/susceptibility example.}  Epidemics spreading in a population for which half have received a leaky vaccine.  Vaccinated individuals are half as infectious and half as susceptible.  We choose $t=0$ to correspond to $1$\% cumulative incidence.  Simulations in a population of $5\times10^5$ individuals (solid) and theory (dashed) are in good agreement.}
\label{fig:heterogeneous}
\end{figure}
One application of this model is to the impact of a partially effective vaccination.   
Vaccination generally reduces the susceptibility of a node, but could either increase or decrease the infectiousness of a node by reducing the severity of symptoms (less sick individuals may shed less virus but also maintain stronger contact intensity while symptomatic).  If only part of the population is vaccinated, then the population can be divided into those who have or have not received vaccination.

Consider a population with a negative binomial degree distribution $\NB(3/2,8/9 )$ with size $r = 3/2$ and probability $p = 8/9$, giving an average degree of $pr/(1-p)=12$ and variance of $pr/(1-p)^2=108$.  For a negative binomial distribution $\NB(r,p)$ we have $\psi(x) = [(1-p)/(1-px)]^r$, so 
\[
\psi(x) =  (9-8x)^{-3/2}
\] 
Assume that half of the population has received a leaky vaccine such that vaccinated nodes have reduced susceptibility, and --- if infected --- reduced infectiousness and infection duration.  Let $\gamma$ be the rate of recovery for unvaccinated nodes and $\beta$ the rate of infection between unvaccinated nodes.  Vaccinated nodes recover at rate $2\gamma$, and the rate of infection between a vaccinated and unvaccinated node (in either direction) is $\beta/2$ while the rate of infection between two vaccinated nodes is $\beta/4$.  The vaccine is distributed uniformly.  Results for $\beta = 0.3$, \ $\gamma = 0.5$ are shown in figure~\ref{fig:heterogeneous}.

\subsection{Populations with assortative mixing by type}
\label{sec:demog}

\begin{figure}
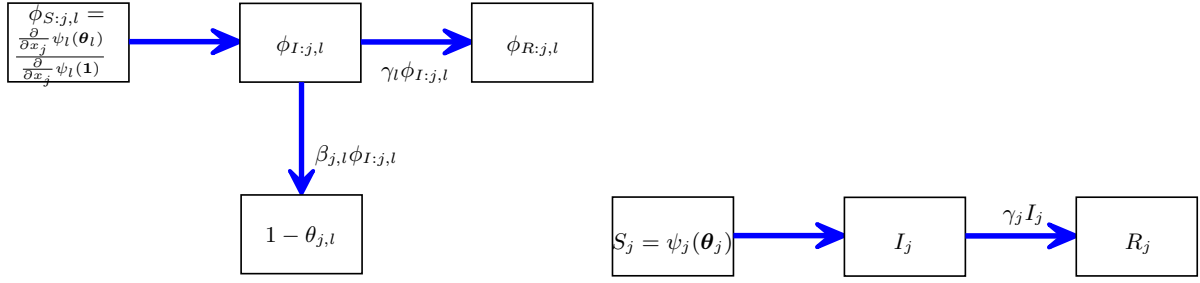

\begin{center}
\scalebox{0.95}{
\begin{picture}(0,0)%
\includegraphics{edgeflux.pdf}%
\end{picture}%
\setlength{\unitlength}{3947sp}%
\begingroup\makeatletter\ifx\SetFigFontNFSS\undefined%
\gdef\SetFigFontNFSS#1#2#3#4#5{%
  \reset@font\fontsize{#1}{#2pt}%
  \fontfamily{#3}\fontseries{#4}\fontshape{#5}%
  \selectfont}%
\fi\endgroup%
\begin{picture}(3883,1833)(2026,-3544)
\put(2149,-2047){\makebox(0,0)[b]{\smash{{\SetFigFontNFSS{9}{9.6}{\familydefault}{\mddefault}{\updefault}{\color[rgb]{0,0,0}\parbox{0.02\textwidth}{~~~$\phi_{S:j,l}=\frac{\pd{}{x_j}\psi_l(\boldsymbol{\theta}_l)}{\pd{}{x_j}\psi_l(\boldsymbol{1})}$}}%
}}}}
\put(3974,-2047){\makebox(0,0)[b]{\smash{{\SetFigFontNFSS{9}{9.6}{\familydefault}{\mddefault}{\updefault}{\color[rgb]{0,0,0}$\phi_{I:j,l}$}%
}}}}
\put(3974,-3310){\makebox(0,0)[b]{\smash{{\SetFigFontNFSS{9}{9.6}{\familydefault}{\mddefault}{\updefault}{\color[rgb]{0,0,0}$1-\theta_{j,l}$}%
}}}}
\put(5511,-2047){\makebox(0,0)[b]{\smash{{\SetFigFontNFSS{9}{9.6}{\familydefault}{\mddefault}{\updefault}{\color[rgb]{0,0,0}$\phi_{R:j,l}$}%
}}}}
\put(4732,-2224){\makebox(0,0)[b]{\smash{{\SetFigFontNFSS{9}{9.6}{\familydefault}{\mddefault}{\updefault}{\color[rgb]{0,0,0}$\gamma_l\phi_{I:j,l}$}%
}}}}
\put(4341,-2774){\makebox(0,0)[b]{\smash{{\SetFigFontNFSS{9}{9.6}{\familydefault}{\mddefault}{\updefault}{\color[rgb]{0,0,0}$\beta_{j,l}\phi_{I:j,l}$}%
}}}}
\end{picture}%
}
\scalebox{0.95}{
\begin{picture}(0,0)%
\includegraphics{standardflux.pdf}%
\end{picture}%
\setlength{\unitlength}{3947sp}%
\begingroup\makeatletter\ifx\SetFigFontNFSS\undefined%
\gdef\SetFigFontNFSS#1#2#3#4#5{%
  \reset@font\fontsize{#1}{#2pt}%
  \fontfamily{#3}\fontseries{#4}\fontshape{#5}%
  \selectfont}%
\fi\endgroup%
\begin{picture}(3883,550)(1651,-1886)
\put(2058,-1700){\makebox(0,0)[b]{\smash{{\SetFigFontNFSS{9}{9.6}{\familydefault}{\mddefault}{\updefault}{\color[rgb]{0,0,0}$S_j = \psi_j(\boldsymbol{\theta}_j)$}%
}}}}
\put(3589,-1700){\makebox(0,0)[b]{\smash{{\SetFigFontNFSS{9}{9.6}{\familydefault}{\mddefault}{\updefault}{\color[rgb]{0,0,0}$I_j$}%
}}}}
\put(5140,-1700){\makebox(0,0)[b]{\smash{{\SetFigFontNFSS{9}{9.6}{\familydefault}{\mddefault}{\updefault}{\color[rgb]{0,0,0}$R_j$}%
}}}}
\put(4382,-1509){\makebox(0,0)[b]{\smash{{\SetFigFontNFSS{9}{9.6}{\familydefault}{\mddefault}{\updefault}{\color[rgb]{0,0,0}$\gamma_j I_j$}%
}}}}
\end{picture}%
}
\end{center}
\caption{\textbf{Assortative mixing by type model.} We separate nodes by type.  We assume that type may influence infectiousness and susceptibility as well as connections.  For simplicity, we assume a finite number of groups.  The resulting system is similar to our result for correlated infectiousness and susceptibility, except in the impact on contacts.}
\label{fig:demog}
\end{figure}
 
In many instances there is biased mixing between or within demographic groups, and the transmission/recovery parameters for the different groups may differ.  For example, the spread of influenza is strongly affected by the increased level of mixing and increased infection rates between children.  Many sexually transmitted diseases are strongly affected by differences in mixing rates and risk behavior among MSM and heterosexual groups.  
For this reason it is useful to have a model accommodating different levels of mixing within and between groups.  The model we derive in this section is equivalent (though simpler) to that of~\cite{volz:bridge}.

Assume that the population is made up of $M$ groups, and let $P_j(k_1,k_2,\ldots,k_M)$ denote the probability a node of group $j$ has $k_l$ contacts with nodes of group $l$ for $l=1,\ldots,M$.  To simplify notation, we denote this by $P_j(\bk)$ where $\bk = (k_1,k_2,\ldots,k_M)$.  We similarly set $\bx = (x_1,x_2,\ldots,x_M)$ and use $\bx^{\bk}$ to denote $x_1^{k_1}x_2^{k_2}\cdots x_M^{k_M}$.  We set
\[
\psi_j(\bx) = \sum_{\bk} P_j(\bk)\bx^{\bk}
\]
and let $\beta_{j,l}$ be the rate of transmission across an edge from group $l$ to group $j$.  We similarly define $\gamma_j$ to be the recovery rate of a group $j$ node.  Define $\theta_{j,l}$ to be the probability an edge to a group $j$ node coming from a group $l$ node has not transmitted infection.   

If our test node $u$ is of type $j$, then the probability that a neighbor $v$ of type $l$ is still susceptible is $\phi_{S:j,l}=\sum_{\bk} k_jP_l(\bk)\theta_{l,1}^{k_1}\theta_{l,2}^{k_2}\cdots \theta_{l,j}^{k_j-1} \cdots\theta_{l,M}^{k_M} / \sum_{\bk} k_j P_l(\bk) = \pd{}{x_j}\psi_l(\btheta_l)/\pd{}{x_j}\psi_l(\boldsymbol{1})$ where $\btheta_l$ denotes the vector $(\theta_{l,1},\theta_{l,2},\ldots,\theta_{l,M})$ and $\boldsymbol{1}$ denotes the vector $(1,1,\ldots,1)$.  We can also show that $v$ has recovered without transmitting to $u$ with probability $\phi_{R:j,l}=\gamma_l(1-\theta_{j,l})/\beta_{j,l}$.  

Figure~\ref{fig:demog} gives
\begin{equation}
\dot{\theta}_{j,l} = -\beta_{j,l} \theta_{j,l} + \beta_{j,l} \frac{\pd{}{x_j}\psi_l(\btheta_l)}{\pd{}{x_j}\psi_l(\boldsymbol{1})}+ \gamma_l(1-\theta_{j,l}) 
\end{equation}
The denominator $\pd{}{x_j}\psi_l(\boldsymbol{1})$ is simply the average of $k_j$ for nodes of group $l$.  We find
\begin{equation}
\dot{R}_j = \gamma_j I_j \qquad \qquad S_j(t) = \psi_j(\btheta_j) \qquad \qquad I_j = 1-S_j - R_j
\end{equation}

As a special case, we can consider a population where the number of contacts a node has with one group is assigned independently of the number that node has with any other group.  We set $P_{j,n}(k)$ to be the probability a node of group $j$ has $k$ edges to a node of group $n$ and define $\psi_{j,n}(x) = \sum_k P_{j,n}(k)x^k$.  Then $\psi_j(\btheta_j)$ factors and may be written as $\psi_{j,1}(\theta_{j,1})\psi_{j,2}(\theta_{j,2})\cdots \psi_{j,M}(\theta_{j,M})$.
In this special case we get
\begin{equation}
\dot{\theta}_{j,l} = - \beta_{j,l} \theta_{j,l} + \beta_{j,l} \frac{\psi_{l,j}'(\theta_{l,j})}{\psi_{l,j}'(1)} \left(\prod_{i \neq j} \psi_{l,i}(\theta_{l,i})\right) + \gamma_l(1-\theta_{j,l})
\label{eqn:demog_indep}
\end{equation}

\subsubsection{Example}
\begin{figure}
\begin{center}
\includegraphics[width=0.48\textwidth]{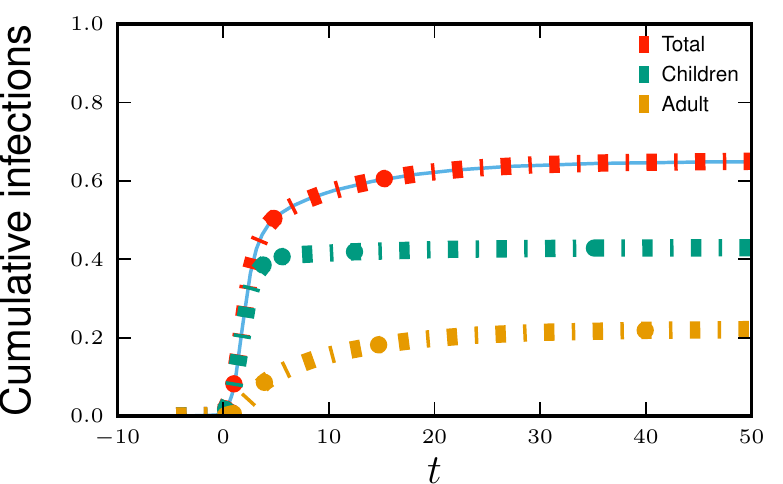}\hfill
\includegraphics[width=0.48\textwidth]{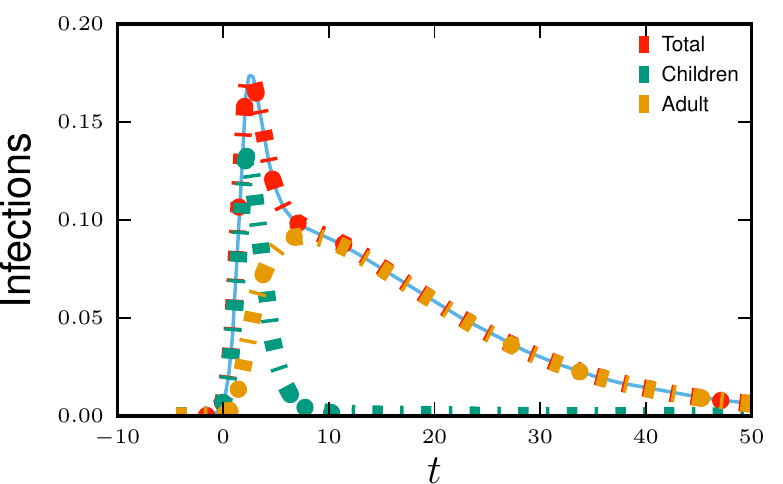}\\
\includegraphics[width=0.48\textwidth]{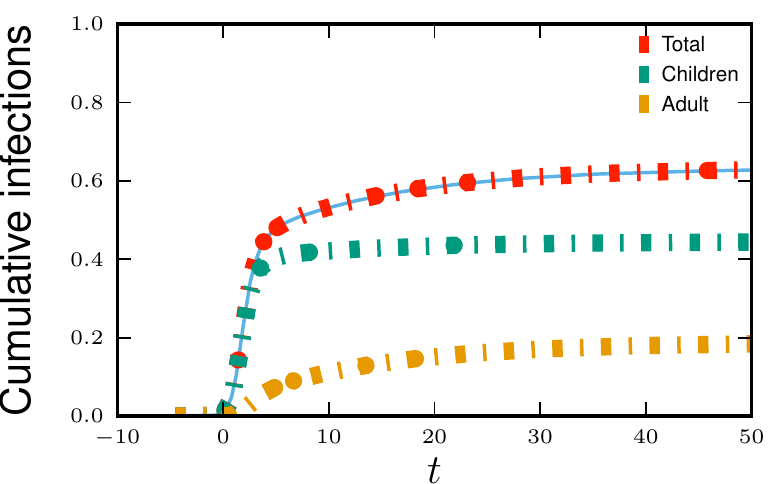}\hfill
\includegraphics[width=0.48\textwidth]{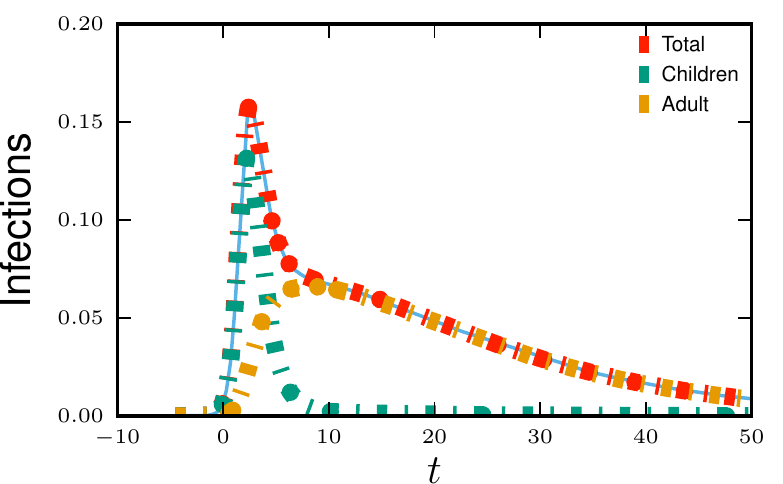}
\end{center}
\caption{\textbf{Assortative mixing by type example.}  Comparison of theory and simulated results for mixing with demographic groups.  We also show the predicted levels of infection in each subgroup.  Simulations in a population of $5\times10^5$ individuals (solid) and theory (dashed) are in good agreement.   The difference between the top and bottom result from changing the correlations of within and between-group mixing. We choose $t=0$ to correspond to $1$\% cumulative incidence.}
\label{fig:biasedmix}
\end{figure}

To demonstrate the ability to capture demographic information, we consider a population made up of two groups, which we arbitrarily label \emph{children} and \emph{adults}.  The between-group degrees are binomially distributed with $\Bi(4,1/2)$, so that the average between-group degree is $2$.  An adult's within-group degree equals its between-group degree.  In contrast, a child's within-group degree is given by $5$ times its between-group degree.  Thus people with higher between-group degree have higher within-group degree.  We get 
\begin{align*}
\psi_a(x_a,x_c) &= (0.5x_ax_c + 0.5)^4\\
\psi_c(x_a,x_c) &= (0.5x_ax_c^5+0.5)^4
\end{align*}
We set the disease parameters to be
\begin{gather*}
\gamma_c = 1\qquad
\gamma_a = 0.1\\
\beta_{c,c} =0.3\qquad
\beta_{c,a} = 0.1\qquad
\beta_{a,a} = 0.1\qquad
\beta_{a,c} =0.1
\end{gather*}
The results are shown in the top of figure~\ref{fig:biasedmix}.

We repeat with the same parameters, but this time with the correlations in degree switched so that higher between-group degrees implies lower within-group degrees.  An adult's within-group degree is $4-\hat{k}$ where $\hat{k}$ is its between-group degree.  A child's within-group degree is $5(4-\hat{k})$ where $\hat{k}$ is its between-group degree.  We have
\begin{align*}
\psi_a(x_a,x_c) &= (0.5x_a+0.5x_c)^4\\
\psi_c(x_a,x_c) &= (0.5x_a+0.5x_c^5)^4
\end{align*}
The results are shown in the bottom of figure~\ref{fig:biasedmix}.

The distribution of within and between-group contacts in the two populations are the same.  The only distinction is that the correlations of within and between-group contacts are different.  This results in differences in the course of the epidemics.  A mass action model could not distinguish between the populations.

In both cases the disease spreads quickly through the child population.  Early on the spread in adults is driven largely by the explosive growth in children.  Because of the correlations of adults' within and between-group degrees, those adults who are infected by children in the first scenario tend to have more adult contacts and infect high-degree adults who in turn infect more high-degree children.  In the second scenario however, infected children tend to infect fewer adults who tend to have fewer adult contacts.  The disease does not grow as quickly, but it also decays less quickly because more high degree nodes remain.

\subsection{Multiple modes of transmission}
\label{sec:multimode}
\begin{figure}
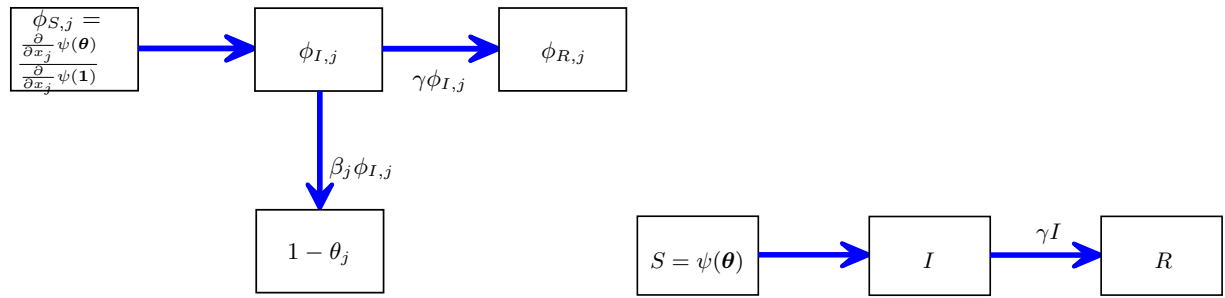

\begin{center}
\begin{picture}(0,0)%
\includegraphics{edgeflux.pdf}%
\end{picture}%
\setlength{\unitlength}{3947sp}%
\begingroup\makeatletter\ifx\SetFigFontNFSS\undefined%
\gdef\SetFigFontNFSS#1#2#3#4#5{%
  \reset@font\fontsize{#1}{#2pt}%
  \fontfamily{#3}\fontseries{#4}\fontshape{#5}%
  \selectfont}%
\fi\endgroup%
\begin{picture}(3883,1833)(2026,-3544)
\put(2149,-2047){\makebox(0,0)[b]{\smash{{\SetFigFontNFSS{9}{9.6}{\familydefault}{\mddefault}{\updefault}{\color[rgb]{0,0,0}\parbox{0.02\textwidth}{~~~$\phi_{S,j}=\frac{\pd{}{x_j}\psi(\boldsymbol{\theta})}{\pd{}{x_j}\psi(\boldsymbol{1})}$}}%
}}}}
\put(3974,-2047){\makebox(0,0)[b]{\smash{{\SetFigFontNFSS{9}{9.6}{\familydefault}{\mddefault}{\updefault}{\color[rgb]{0,0,0}$\phi_{I,j}$}%
}}}}
\put(3974,-3310){\makebox(0,0)[b]{\smash{{\SetFigFontNFSS{9}{9.6}{\familydefault}{\mddefault}{\updefault}{\color[rgb]{0,0,0}$1-\theta_j$}%
}}}}
\put(5511,-2047){\makebox(0,0)[b]{\smash{{\SetFigFontNFSS{9}{9.6}{\familydefault}{\mddefault}{\updefault}{\color[rgb]{0,0,0}$\phi_{R,j}$}%
}}}}
\put(4732,-2224){\makebox(0,0)[b]{\smash{{\SetFigFontNFSS{9}{9.6}{\familydefault}{\mddefault}{\updefault}{\color[rgb]{0,0,0}$\gamma\phi_{I,j}$}%
}}}}
\put(4241,-2774){\makebox(0,0)[b]{\smash{{\SetFigFontNFSS{9}{9.6}{\familydefault}{\mddefault}{\updefault}{\color[rgb]{0,0,0}$\beta_j\phi_{I,j}$}%
}}}}
\end{picture}%
\scalebox{0.95}{
\begin{picture}(0,0)%
\includegraphics{standardflux.pdf}%
\end{picture}%
\setlength{\unitlength}{3947sp}%
\begingroup\makeatletter\ifx\SetFigFontNFSS\undefined%
\gdef\SetFigFontNFSS#1#2#3#4#5{%
  \reset@font\fontsize{#1}{#2pt}%
  \fontfamily{#3}\fontseries{#4}\fontshape{#5}%
  \selectfont}%
\fi\endgroup%
\begin{picture}(3883,550)(1651,-1886)
\put(2058,-1700){\makebox(0,0)[b]{\smash{{\SetFigFontNFSS{9}{9.6}{\familydefault}{\mddefault}{\updefault}{\color[rgb]{0,0,0}$S = \psi(\boldsymbol{\theta})$}%
}}}}
\put(3589,-1700){\makebox(0,0)[b]{\smash{{\SetFigFontNFSS{9}{9.6}{\familydefault}{\mddefault}{\updefault}{\color[rgb]{0,0,0}$I$}%
}}}}
\put(5140,-1700){\makebox(0,0)[b]{\smash{{\SetFigFontNFSS{9}{9.6}{\familydefault}{\mddefault}{\updefault}{\color[rgb]{0,0,0}$R$}%
}}}}
\put(4382,-1509){\makebox(0,0)[b]{\smash{{\SetFigFontNFSS{9}{9.6}{\familydefault}{\mddefault}{\updefault}{\color[rgb]{0,0,0}$\gamma I$}%
}}}}
\end{picture}%
}
\end{center}
\caption{\textbf{Multiple modes of transmission model.}  Flow diagram showing the flux of edges for the $j$-th contact type for a disease which has multiple modes of transmission.}
\label{fig:modes}
\end{figure}
 
Rather than having different types of nodes, there may be multiple modes of transmission with different mixing and infection rates for each mode.  For example, HIV may spread through sexual contact and needle-sharing.  The sexual contact network may have little to no relation to the needle-sharing network.  Assume there are $M$ types of contacts and that the rate of transmission along contacts of mode $j$ is given by $\beta_j$.  Let the joint distribution of the number of each mode of contacts be given by $P(\bk)$ where $\bk=(k_1,k_2,\ldots,k_M)$.  Assume recovery rates are independent of how infection was acquired.  We set $\bx = (x_1,\ldots,x_M)$ and denote $x_1^{k_1}\cdots x_M^{k_M}$ by $\bx^{\bk}$.  We define
\[
\psi(\bx)= \sum_{\bk} P(\bk) \bx^{\bk}
\]
We can apply the same method to each edge type as shown in figure~\ref{fig:modes}.  We set $\phi_{S,j}$ to be the probability that an edge of type $j$ connects the test node to a susceptible neighbor.  If $\theta_j$ is the probability a contact of type $j$ has not yet transmitted infection, we set $\btheta=(\theta_1,\theta_2,\ldots,\theta_M)$.   We find $\phi_{S,j} = \pd{}{x_j} \psi(\btheta)/\pd{}{x_j}\psi(\boldsymbol{1})$ and similarly the probability an edge of type $j$ connects to a recovered neighbor which did not transmit is $\phi_{R,j}=\gamma(1-\theta_j)/\beta_j$.  As before we find $S(t) = \psi(\btheta)$.  We have
\begin{align}
\dot{\theta}_j &= - \beta_j \theta_j + \beta_j \frac{\pd{}{x_j}\psi(\btheta)}{\pd{}{x_j}\psi(\boldsymbol{1})} + \gamma (1-\theta_j) \label{eqn:multi_mode_thetadot}\\
\dot{R} &= \gamma I \qquad\qquad S =\psi(\btheta) \qquad\qquad I = 1-S-R
\end{align}
In the case where the degree with respect to one contact type is independent of that with respect to another, we can simplify equation~\eqref{eqn:multi_mode_thetadot} to be similar to equation~\eqref{eqn:demog_indep}.

\subsubsection{Example}
\begin{figure}
\begin{center}
\includegraphics[width=0.48\textwidth]{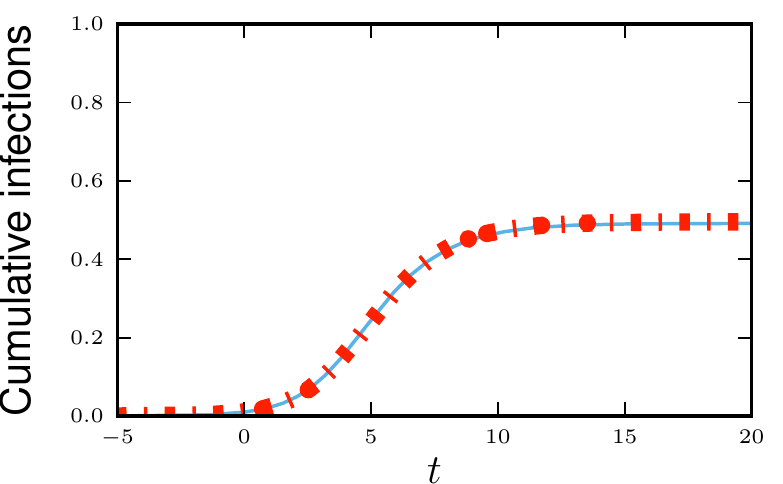}\hfill
\includegraphics[width=0.48\textwidth]{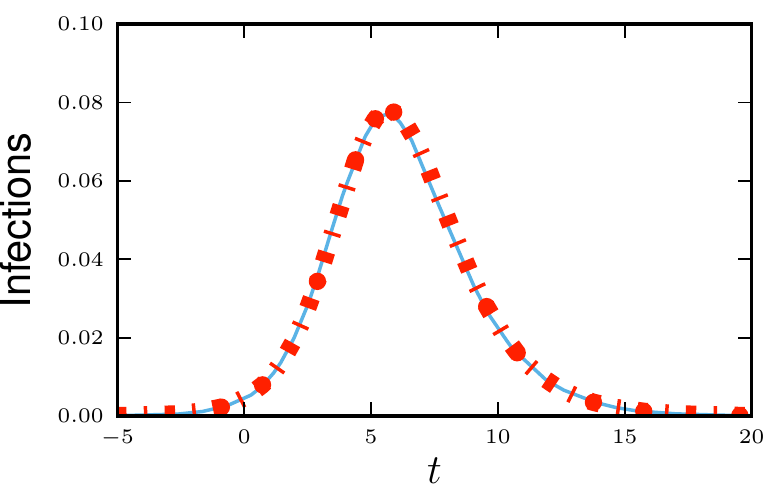}
\end{center}
\caption{\textbf{Multiple  modes of transmission example.} Disease spread in a population with three different types of contacts, each with a different degree distribution.  Simulations in a population of $5\times10^5$ individuals (solid) and theory (dashed) are in good agreement.  We choose $t=0$ to correspond to $1$\% cumulative incidence.} 
\label{fig:multi_mode}
\end{figure}

We consider a population with three different types of contacts.  We take $k_1$, $k_2$, and $k_3$ to denote the number of each type of contact a node has.  We assume that contact type $1$ is binomially distributed with $\Bi(2,0.5)$ (giving a mean of $1$).  Contact type $2$ is geometrically distributed, with mean $2$.  Contact type $3$ has a negative binomial distribution $\NB(1,1/4)$ with mean $1/3$ and variance $4/9$.  The numbers of contacts an individual has of each type are assigned independently.
We find
\[
\psi(x_1,x_2,x_3) = \frac{(x_1+1)^2}{4} \frac{x_2}{2-x_2} \frac{3}{4-x_3}
\]
We take $\gamma = 1$ and set $\beta_1=1$, \ $\beta_2=0.5$, and $\beta_3 = 3$ for each contact type.
We compare simulation and theory in figure~\ref{fig:multi_mode}.

\subsection{Multiple infectious stages}
\label{sec:multistage}
\begin{figure}
\begin{center}
\begin{picture}(0,0)%
\includegraphics{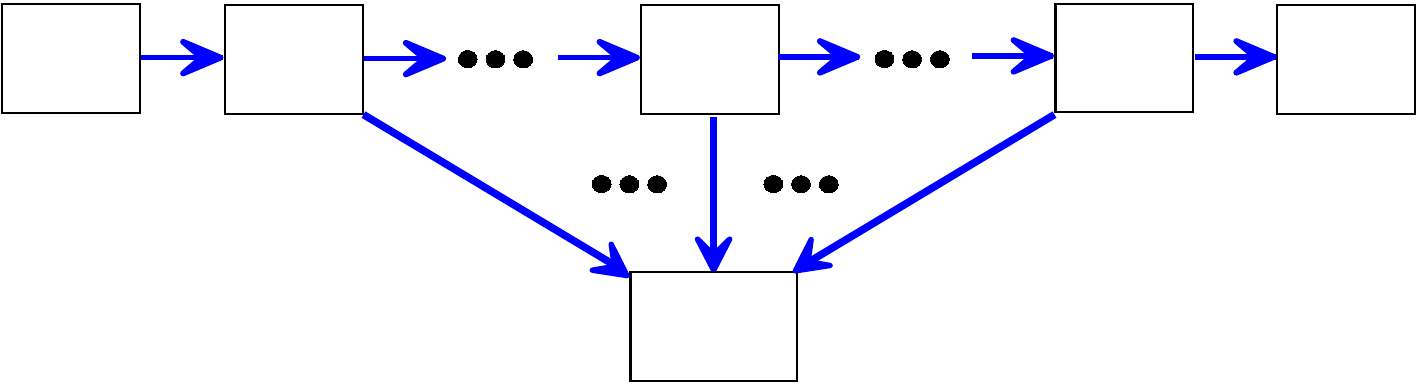}%
\end{picture}%
\setlength{\unitlength}{3947sp}%
\begingroup\makeatletter\ifx\SetFigFontNFSS\undefined%
\gdef\SetFigFontNFSS#1#2#3#4#5{%
  \reset@font\fontsize{#1}{#2pt}%
  \fontfamily{#3}\fontseries{#4}\fontshape{#5}%
  \selectfont}%
\fi\endgroup%
\begin{picture}(6800,1850)(2701,-3036)
\put(3037,-1530){\makebox(0,0)[b]{\smash{{\SetFigFontNFSS{9}{9.6}{\familydefault}{\mddefault}{\updefault}{\color[rgb]{0,0,0}$\phi_S=\frac{\psi'(\theta)}{\psi'(1)}$}%
}}}}
\put(4114,-1530){\makebox(0,0)[b]{\smash{{\SetFigFontNFSS{9}{9.6}{\familydefault}{\mddefault}{\updefault}{\color[rgb]{0,0,0}$\phi_{I,1}$}%
}}}}
\put(6120,-1530){\makebox(0,0)[b]{\smash{{\SetFigFontNFSS{9}{9.6}{\familydefault}{\mddefault}{\updefault}{\color[rgb]{0,0,0}$\phi_{I,j}$}%
}}}}
\put(8101,-1530){\makebox(0,0)[b]{\smash{{\SetFigFontNFSS{9}{9.6}{\familydefault}{\mddefault}{\updefault}{\color[rgb]{0,0,0}$\phi_{I,M}$}%
}}}}
\put(9169,-1530){\makebox(0,0)[b]{\smash{{\SetFigFontNFSS{9}{9.6}{\familydefault}{\mddefault}{\updefault}{\color[rgb]{0,0,0}$\phi_R$}%
}}}}
\put(6120,-2826){\makebox(0,0)[b]{\smash{{\SetFigFontNFSS{9}{9.6}{\familydefault}{\mddefault}{\updefault}{\color[rgb]{0,0,0}$1-\theta$}%
}}}}
\put(4804,-1704){\makebox(0,0)[b]{\smash{{\SetFigFontNFSS{9}{9.6}{\familydefault}{\mddefault}{\updefault}{\color[rgb]{0,0,0}$\gamma_1\phi_{I,1}$}%
}}}}
\put(5388,-1344){\makebox(0,0)[b]{\smash{{\SetFigFontNFSS{9}{9.6}{\familydefault}{\mddefault}{\updefault}{\color[rgb]{0,0,0}$\gamma_{j-1}\phi_{I,j-1}$}%
}}}}
\put(6716,-1684){\makebox(0,0)[b]{\smash{{\SetFigFontNFSS{9}{9.6}{\familydefault}{\mddefault}{\updefault}{\color[rgb]{0,0,0}$\gamma_j\phi_{I,j}$}%
}}}}
\put(7306,-1344){\makebox(0,0)[b]{\smash{{\SetFigFontNFSS{9}{9.6}{\familydefault}{\mddefault}{\updefault}{\color[rgb]{0,0,0} $\gamma_{M-1}\phi_{I,M-1}$}%
}}}}
\put(8648,-1844){\makebox(0,0)[b]{\smash{{\SetFigFontNFSS{9}{9.6}{\familydefault}{\mddefault}{\updefault}{\color[rgb]{0,0,0}$\gamma_M\phi_{I,M}$}%
}}}}
\put(5281,-2520){\makebox(0,0)[b]{\smash{{\SetFigFontNFSS{9}{9.6}{\familydefault}{\mddefault}{\updefault}{\color[rgb]{0,0,0}$\beta_1\phi_{I,1}$}%
}}}}
\put(7016,-2520){\makebox(0,0)[b]{\smash{{\SetFigFontNFSS{9}{9.6}{\familydefault}{\mddefault}{\updefault}{\color[rgb]{0,0,0}$\beta_M\phi_{I,M}$}%
}}}}
\put(6397,-2241){\makebox(0,0)[b]{\smash{{\SetFigFontNFSS{9}{9.6}{\familydefault}{\mddefault}{\updefault}{\color[rgb]{0,0,0}$\beta_j\phi_{I,j}$}%
}}}}
\end{picture}\\[30pt]%
\begin{picture}(0,0)%
\includegraphics{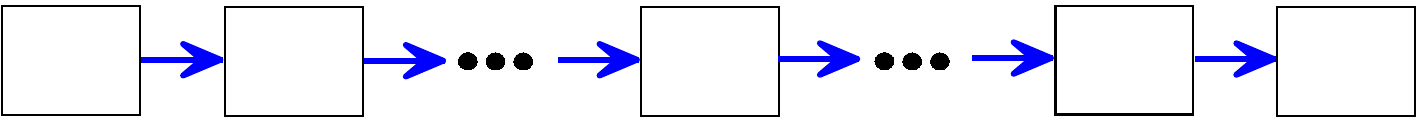}%
\end{picture}%
\setlength{\unitlength}{3947sp}%
\begingroup\makeatletter\ifx\SetFigFontNFSS\undefined%
\gdef\SetFigFontNFSS#1#2#3#4#5{%
  \reset@font\fontsize{#1}{#2pt}%
  \fontfamily{#3}\fontseries{#4}\fontshape{#5}%
  \selectfont}%
\fi\endgroup%
\begin{picture}(6800,573)(2701,-1759)
\put(3037,-1530){\makebox(0,0)[b]{\smash{{\SetFigFontNFSS{9}{9.6}{\familydefault}{\mddefault}{\updefault}{\color[rgb]{0,0,0}$S=\psi(\theta)$}%
}}}}
\put(4114,-1530){\makebox(0,0)[b]{\smash{{\SetFigFontNFSS{9}{9.6}{\familydefault}{\mddefault}{\updefault}{\color[rgb]{0,0,0}$I_1$}%
}}}}
\put(6120,-1530){\makebox(0,0)[b]{\smash{{\SetFigFontNFSS{9}{9.6}{\familydefault}{\mddefault}{\updefault}{\color[rgb]{0,0,0}$I_j$}%
}}}}
\put(8101,-1530){\makebox(0,0)[b]{\smash{{\SetFigFontNFSS{9}{9.6}{\familydefault}{\mddefault}{\updefault}{\color[rgb]{0,0,0}$I_M$}%
}}}}
\put(9169,-1530){\makebox(0,0)[b]{\smash{{\SetFigFontNFSS{9}{9.6}{\familydefault}{\mddefault}{\updefault}{\color[rgb]{0,0,0}$R$}%
}}}}
\put(4834,-1744){\makebox(0,0)[b]{\smash{{\SetFigFontNFSS{9}{9.6}{\familydefault}{\mddefault}{\updefault}{\color[rgb]{0,0,0}$\gamma_1I_1$}%
}}}}
\put(5508,-1744){\makebox(0,0)[b]{\smash{{\SetFigFontNFSS{9}{9.6}{\familydefault}{\mddefault}{\updefault}{\color[rgb]{0,0,0}$\gamma_{j-1}I_{j-1}$}%
}}}}
\put(6646,-1744){\makebox(0,0)[b]{\smash{{\SetFigFontNFSS{9}{9.6}{\familydefault}{\mddefault}{\updefault}{\color[rgb]{0,0,0}$\gamma_jI_j$}%
}}}}
\put(7406,-1744){\makebox(0,0)[b]{\smash{{\SetFigFontNFSS{9}{9.6}{\familydefault}{\mddefault}{\updefault}{\color[rgb]{0,0,0}$\gamma_{M-1}I_{M-1}$}%
}}}}
\put(8648,-1744){\makebox(0,0)[b]{\smash{{\SetFigFontNFSS{9}{9.6}{\familydefault}{\mddefault}{\updefault}{\color[rgb]{0,0,0}$\gamma_MI_M$}%
}}}}
\end{picture}%
\end{center}
\caption{\textbf{Multiple infectious stages model.}  Flow diagram for a disease with several infected stages.  When a disease progresses through several states (or has an infectious period that is not exponentially distributed) it is convenient to use a stage-progression model to represent the state of an edge.}
\label{fig:stages}
\end{figure}
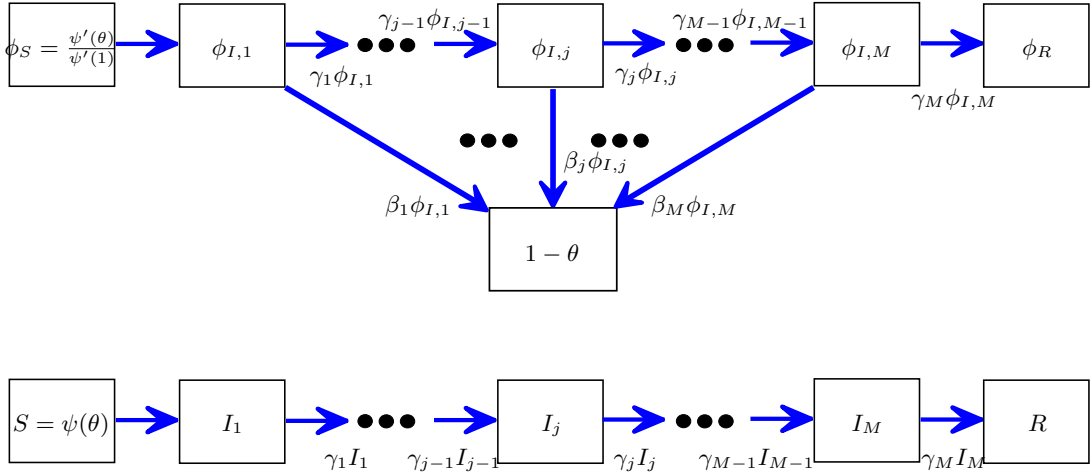

There are a number of diseases with multiple infectious stages such as Tuberculosis and HIV.  Some diseases begin with a non-infectious latent phase, some begin with a highly infectious acute stage before settling into a long-term chronic stage, and others oscillate between phases of high and low infectiousness.  To model such situations we adapt a standard chain progression model, for which there are $M$ infectious phases shown in figure~\ref{fig:stages}.  We are not able to explicitly solve for all variables in terms of $\theta$, so we must find the fluxes between the compartments.  We can still find $\phi_S = \psi'(\theta)/\psi'(1)$, so we are able to find $\phi_{I,1}$ in terms of the other variables using $\phi_S + \phi_R + \sum_j \phi_{I,j} = \theta$.  We obtain
\begin{align}
\dot{\theta} &= -\sum \beta_j \phi_{I,j}\\
\phi_{I,1} &= \theta - \frac{\psi'(\theta)}{\psi'(1)} - \phi_R - \sum_{j=2}^M \phi_{I,j}\\
\dot{\phi}_{I,j} &= \gamma_{j-1}\phi_{I,j-1} - \gamma_j \phi_{I,j} -\beta_j\phi_{I,j}\qquad\qquad M\geq j>1\\
\dot{\phi}_R &= \gamma_M\phi_{I,M}\\
\dot{I}_j &= \gamma_{j-1}I_{j-1} - \gamma_j I_j\qquad\qquad M\geq j>1\\
\dot{R} &= \gamma_{M} I_M \qquad\qquad S = \psi(\theta) \qquad\qquad
I_1 = 1- S - R - \sum_{j=2}^M I_j
\end{align}
Where $I_j$ is the proportion of the population in the $j$-th infectious class.

\subsubsection{Example}
\label{sec:multi_stage_ex}
\begin{figure}
  \begin{center}
  \includegraphics[width=0.48\textwidth]{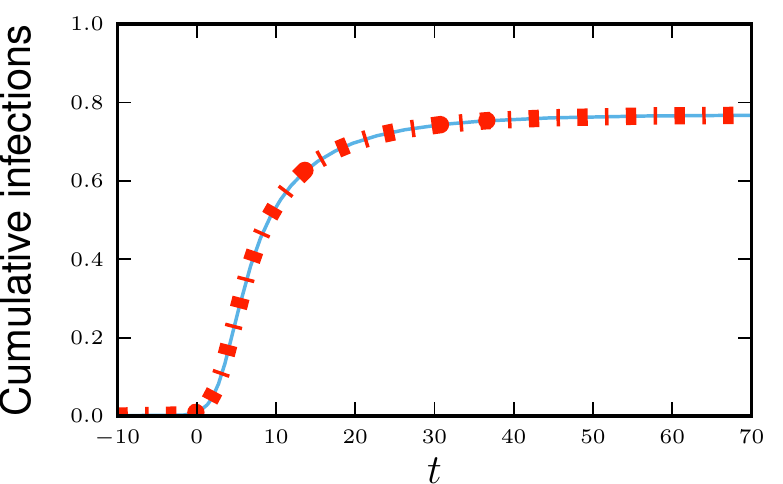}\hfill
  \includegraphics[width=0.48\textwidth]{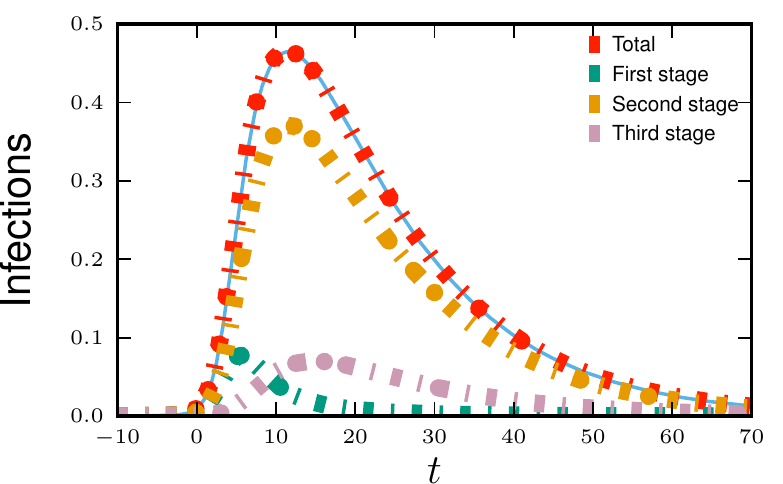}
 \end{center}
  \caption{\textbf{Multiple infectious stages example.}  The spread of the disease described in~\ref{sec:multi_stage_ex} with three infectious stages.   Simulations in a population of $5\times10^5$ individuals (solid) and theory (dashed) are in good agreement.  We choose $t=0$ to correspond to $1$\% cumulative incidence.}
  \label{fig:multistage}
\end{figure}

Consider now the spread of a disease for which there are three infectious stages.  The first stage is moderately infectious but not long, with $\beta_1 = 0.2$, \ $\gamma_1=1$.  The second stage is much longer, but has a substantially lower infectiousness, with $\beta_2 = 0.01$ and $\gamma_2 = 0.08$.  The final stage has an intermediate duration but substantially higher infectiousness, with $\beta_3 = 2$ and $\gamma_3 = 0.4$.  We assume that the disease spreads in a population with degree distribution $\NB(1,4/5)$ having mean $4$ and variance $20$ with
\[
\psi(x)=(5-4x)^{-1}
\]
The results are shown in figure~\ref{fig:multistage}.

\section{Dynamic Networks and Serosorting}
For some diseases, it is not uncommon for individuals to actively seek out contacts of similar disease status (as in HIV~\cite{parsons:serosort} or Leprosy~\cite{leviticus13}) or even discordant status (as in ``chicken pox parties'' or ``swine flu parties'').  
This is commonly known as serosorting.  To study these populations, we must use dynamic network models, which we developed in~\cite{miller:ebcm_overview}.

We study serosorting in two models.  In the first, we use an ``actual degree'' model where an individual has a given number of potential contacts, of which only a proportion are active at any given time.  In the second, we use an ``expected degree'' model in which individuals break any existing contacts at a fixed rate, but different individuals may create contacts at differing rates leading to a variation in the expected number of contacts across the population.

For simplicity, we will assume that there is no recovered class, and once infected an individual remains infected.  This restriction is easily removed, but by using it, we are able to simplify the model and reduce the number of parameters needed.  We again consider a test node, and as before we assume that if infected the test node does not cause any infections.  We make an additional assumption that if infected the test node continues to behave as if it were susceptible, and that its potential neighbors treat it as if it were susceptible.  We can think of the test node as an individual who is immune, but is unaware of that immunity, and we track the probability that the test node has not yet received a sufficient dose to infect a non-immune individual.

\subsection{Actual Degree Serosorting model}
\label{sec:fd_serosort}

In the actual degree formulation, we think of an individual as having $k$ \emph{stubs} or half-edges. These stubs may be active (and connected to another node's stub) or dormant (and available to form new edges).

When an edge breaks, the corresponding stubs enter a dormant phase.  We assume that the rate a dormant edge belonging to a susceptible individual finds a new susceptible neighbor is $\eta_{1,SS}$, and the rate it finds a new infected neighbor is $\eta_{1,SI}$.  These may depend on the density of susceptible and infected individuals in the population.  Similarly, active edges break at rates depending on the status of the nodes.  Edges between susceptible nodes break at rate $\eta_{2,SS}$, edges between a susceptible and infected node breaks at rate $\eta_{2,SI}$, and edges between infected nodes break at rate $\eta_{2,II}$.

\begin{figure}
(a) \begin{picture}(0,0)%
\includegraphics{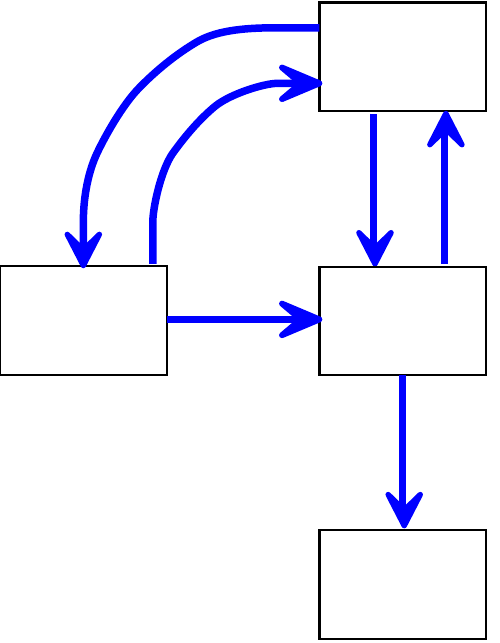}%
\end{picture}%
\setlength{\unitlength}{3947sp}%
\begingroup\makeatletter\ifx\SetFigFontNFSS\undefined%
\gdef\SetFigFontNFSS#1#2#3#4#5{%
  \reset@font\fontsize{#1}{#2pt}%
  \fontfamily{#3}\fontseries{#4}\fontshape{#5}%
  \selectfont}%
\fi\endgroup%
\begin{picture}(2396,3083)(2106,-4024)
\put(2487,-2536){\makebox(0,0)[b]{\smash{{\SetFigFontNFSS{8}{9.6}{\familydefault}{\mddefault}{\updefault}{\color[rgb]{0,0,0}$\phi_S$}%
}}}}
\put(4048,-2536){\makebox(0,0)[b]{\smash{{\SetFigFontNFSS{8}{9.6}{\familydefault}{\mddefault}{\updefault}{\color[rgb]{0,0,0}$\phi_I$}%
}}}}
\put(4048,-1260){\makebox(0,0)[b]{\smash{{\SetFigFontNFSS{8}{9.6}{\familydefault}{\mddefault}{\updefault}{\color[rgb]{0,0,0}$\phi_D$}%
}}}}
\put(4048,-3812){\makebox(0,0)[b]{\smash{{\SetFigFontNFSS{8}{9.6}{\familydefault}{\mddefault}{\updefault}{\color[rgb]{0,0,0}$1-\theta$}%
}}}}
\put(3303,-2848){\makebox(0,0)[b]{\smash{{\SetFigFontNFSS{8}{9.6}{\familydefault}{\mddefault}{\updefault}{\color[rgb]{0,0,0}$\beta\phi_I\phi_S\frac{\psi''(\theta)}{\psi'(\theta)}$}%
}}}}
\put(4395,-1937){\makebox(0,0)[lb]{\smash{{\SetFigFontNFSS{8}{9.6}{\familydefault}{\mddefault}{\updefault}{\color[rgb]{0,0,0}$\eta_{2,SI}\phi_I$}%
}}}}
\put(4182,-3118){\makebox(0,0)[lb]{\smash{{\SetFigFontNFSS{8}{9.6}{\familydefault}{\mddefault}{\updefault}{\color[rgb]{0,0,0}$\beta\phi_I$}%
}}}}
\put(3037,-1780){\makebox(0,0)[lb]{\smash{{\SetFigFontNFSS{8}{9.6}{\familydefault}{\mddefault}{\updefault}{\color[rgb]{0,0,0}$\eta_{2,SS}\phi_S$}%
}}}}
\put(2615,-1310){\makebox(0,0)[rb]{\smash{{\SetFigFontNFSS{8}{9.6}{\familydefault}{\mddefault}{\updefault}{\color[rgb]{0,0,0}$\eta_{1,SS}\phi_D$}%
}}}}
\put(3827,-1919){\makebox(0,0)[rb]{\smash{{\SetFigFontNFSS{8}{9.6}{\familydefault}{\mddefault}{\updefault}{\color[rgb]{0,0,0}$\eta_{1,SI}\phi_D$}%
}}}}
\end{picture}%
\hfill 
(b) \begin{picture}(0,0)%
\includegraphics{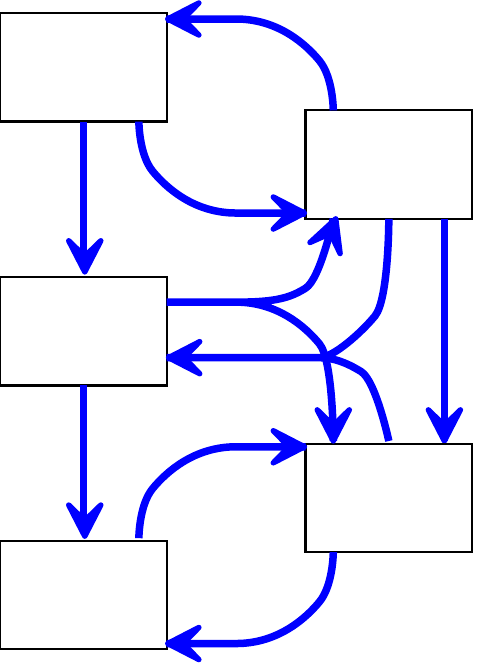}%
\end{picture}%
\setlength{\unitlength}{3947sp}%
\begingroup\makeatletter\ifx\SetFigFontNFSS\undefined%
\gdef\SetFigFontNFSS#1#2#3#4#5{%
  \reset@font\fontsize{#1}{#2pt}%
  \fontfamily{#3}\fontseries{#4}\fontshape{#5}%
  \selectfont}%
\fi\endgroup%
\begin{picture}(2392,3183)(1601,-3306)
\put(1988,-499){\makebox(0,0)[b]{\smash{{\SetFigFontNFSS{8}{9.6}{\familydefault}{\mddefault}{\updefault}{\color[rgb]{0,0,0}$M_{SS}$}%
}}}}
\put(1988,-1764){\makebox(0,0)[b]{\smash{{\SetFigFontNFSS{8}{9.6}{\familydefault}{\mddefault}{\updefault}{\color[rgb]{0,0,0}$M_{SI}$}%
}}}}
\put(1988,-3024){\makebox(0,0)[b]{\smash{{\SetFigFontNFSS{8}{9.6}{\familydefault}{\mddefault}{\updefault}{\color[rgb]{0,0,0}$M_{II}$}%
}}}}
\put(3454,-2559){\makebox(0,0)[b]{\smash{{\SetFigFontNFSS{8}{9.6}{\familydefault}{\mddefault}{\updefault}{\color[rgb]{0,0,0}$\pi_I$}%
}}}}
\put(3454,-970){\makebox(0,0)[b]{\smash{{\SetFigFontNFSS{8}{9.6}{\familydefault}{\mddefault}{\updefault}{\color[rgb]{0,0,0}$\pi_S$}%
}}}}
\put(3525,-1725){\makebox(0,0)[b]{\smash{{\SetFigFontNFSS{8}{9.6}{\familydefault}{\mddefault}{\updefault}{\color[rgb]{0,0,0}\rotatebox{270}{$\eta_{1,SI}\pi_S$}}%
}}}}
\put(3004,-1708){\makebox(0,0)[b]{\smash{{\SetFigFontNFSS{8}{9.6}{\familydefault}{\mddefault}{\updefault}{\color[rgb]{0,0,0}\rotatebox{315}{$\frac{\eta_{2,SI}M_{SI}}{2}$}}%
}}}}
\put(3125,-3116){\makebox(0,0)[lb]{\smash{{\SetFigFontNFSS{8}{9.6}{\familydefault}{\mddefault}{\updefault}{\color[rgb]{0,0,0}$\eta_{1,II}\pi_I$}%
}}}}
\put(3786,-1309){\makebox(0,0)[lb]{\smash{{\SetFigFontNFSS{8}{9.6}{\familydefault}{\mddefault}{\updefault}{\color[rgb]{0,0,0}\rotatebox{270}{$\beta\phi_D\phi_I\frac{\psi''(\theta)}{\psi'(1)}$}}%
}}}}
\put(2343,-2209){\makebox(0,0)[lb]{\smash{{\SetFigFontNFSS{8}{9.6}{\familydefault}{\mddefault}{\updefault}{\color[rgb]{0,0,0}$\eta_{2,II}M_{II}$}%
}}}}
\end{picture}%
\caption{\textbf{Fixed-degree serosorting model.} Flow diagram showing the interplay involved in serosorting with fixed-degree.  We do not consider a recovered class, which simplifies the equations.  The framework can be adapted to include a recovered class.  The $M$ variables represent the total proportion of stubs involved in edges between the two types and the $\pi$ variables are the proportion of dormant stubs belonging to nodes of each type.  The $\phi$ variables are as before.  For the right hand side, we are able to determine most of the variables analytically, so we only need the fluxes into and out of $\pi_I$.  We expect that the edge breaking and rejoining rates $\eta$ will depend on values of $\pi_S$ and $\pi_I$.}
\label{fig:fixdeg_sero}
\end{figure}
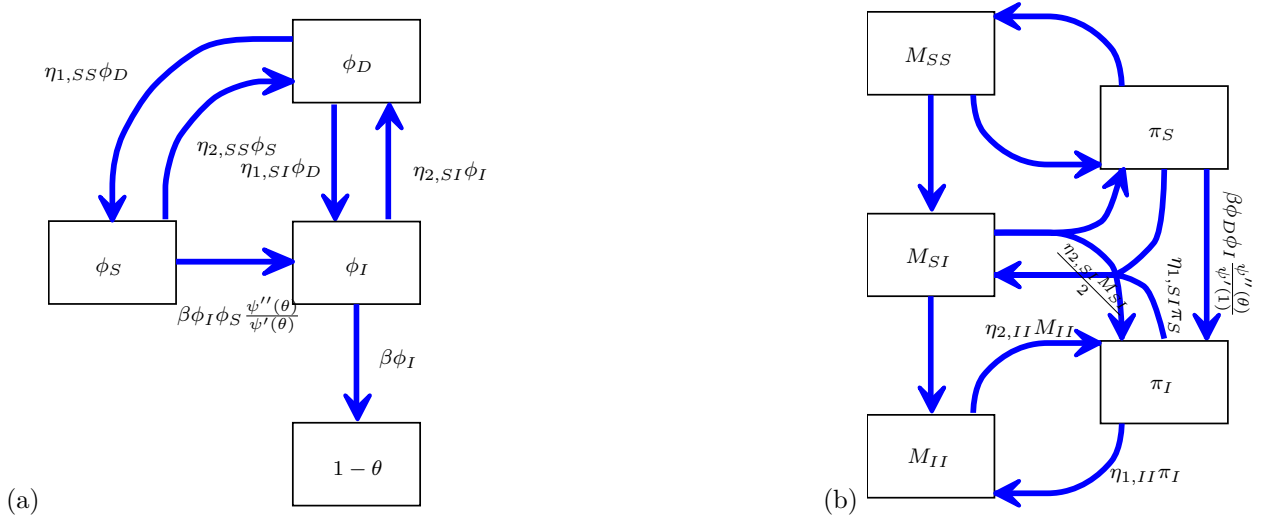

We define $\phi_S$, $\phi_I$, $\phi_R$, and $\phi_D$ to be the probability that a stub belonging to the test node $u$ has never been part of an edge that transmitted infection to $u$, and that the stub is currently connected to a susceptible, infected, or recovered node or is dormant respectively.  The fluxes between these states are shoen in figure~\ref{fig:fixdeg_sero}.  Unlike in previous cases, we are unable to explicitly calculate $\phi_S$, so we must track the flux into and out of $\phi_S$.  The fluxes between $\phi_S$ and $\phi_D$ are straightforward.  However, the flux from $\phi_S$ to $\phi_I$ requires more attention.  We repeat our derivation from~\cite{miller:ebcm_overview}.  Consider a neighbor $v$ of the test node $u$ having the following properties: the stub belonging to $u$ never transmitted to $u$ prior to joining with the stub belonging to $v$, and similarly the stub belonging to $v$ never transmitted to $v$ prior to joining with the stub belonging to $u$.
Given this, the probability $v$ is susceptible is $q=\sum_k k P(k)\theta^{k-1}/\ave{K}= \psi'(\theta)/\psi'(1)$.  Thus, given that $v$ is susceptible, the rate $v$ becomes infected is 
\[
-\frac{\dot{q}}{q} = -\frac{\dot{\theta}\psi''(\theta)/\psi'(1)}{\psi'(\theta)/\psi'(1)}=\beta \phi_I \frac{\psi''(\theta)}{\psi'(\theta)}
\]   
Thus the flux from $\phi_S$ to $\phi_I$ is  $\beta \phi_S \phi_I \psi''(\theta)/\psi'(\theta)$, the product of $\phi_S$, the probability that a stub has not transmitted infection to the test node and connects to a susceptible node, with  $\beta \phi_I \psi''(\theta)/\psi'(\theta)$, the rate that the neighbor becomes infected given that the stub has not transmitted and connects to a susceptible node.  

We need to account for the number of stubs that are in edges between different classes of nodes or are dormant.  We use $M_{SS}$ to be the proportion of all stubs that are in edges between susceptible nodes.  Equivalently this is the probability that a stub is active, connects to a susceptible node, and belongs to a susceptible node: $M_{SS} = \phi_S \psi'(\theta)/\psi'(1)$.  We similarly define $M_{SI}$ to be the proportion of all stubs that are in edges between a susceptible and infected node.  We calculate this by finding the probability a stub is active, connects to an infected node, but belongs to a susceptible node but we must multiply by $2$ because this only counts one stub in each edge.  We get $M_{SI} = 2\phi_I \psi'(\theta)/\psi'(1)$ .  We also define $M_{II}$ to be the proportion of all stubs in edges between infected nodes.  We will calculate its value later.  We set $\pi_S$ to be the proportion of stubs that are dormant and belong to susceptible nodes and $\pi_I$ to be the proportion of stubs that are dormant and belong to infected nodes.  The value of $\pi_S$ can be calculated as for the dormant contact case to be $\pi_S = \phi_D \psi'(\theta)/\psi'(1)$.  The value of $\pi_I$ is calculated by finding the fluxes out of the other states.  Once we have all of these variables, we have $M_{II} = 1 - M_{SS} - M_{SI} - \pi_S - \pi_I$.

Following figure~\ref{fig:fixdeg_sero} we find
\begin{align}
\dot{\theta} &= -\beta \phi_I\\
\dot{\phi}_S &= \eta_{1,SS}(\pi_S,\pi_I)\phi_D - \eta_{2,SS}(\pi_S,\pi_I)\phi_S - \beta \phi_I\phi_S\frac{\psi''(\theta)}{\psi'(\theta)}\\
\dot{\phi}_I &= \eta_{1,SI}(\pi_S,\pi_I)\phi_D - \eta_{2,SI}(\pi_S,\pi_I)\phi_I - \beta \phi_I + \beta \phi_I\phi_S\frac{\psi''(\theta)}{\psi'(\theta)}\\
\dot{\phi}_D &= \eta_{2,SS}(\pi_S,\pi_I)\phi_S + \eta_{2,SI}(\pi_S,\pi_I)\phi_I - [\eta_{1,SS}(\pi_S,\pi_I)+\eta_{1,SI}(\pi_S,\pi_I)]\phi_D\\
\pi_S &= \phi_D \frac{\psi'(\theta)}{\psi'(1)}\\
\dot{\pi}_I &= \frac{\eta_{2,SI}(\pi_S,\pi_I)}{2}M_{SI} - \pi_S \eta_{1,SI}(\pi_S,\pi_I) + \eta_{2,II}(\pi_S,\pi_I)M_{II}-\pi_I\eta_{1,II}+\beta \phi_D \psi_I\frac{\psi''(\theta)}{\psi'(1)}\\
M_{SS} &= \phi_S \frac{\psi'(\theta)}{\psi'(1)}\\
M_{SI} &= 2 \phi_I \frac{\psi'(\theta)}{\psi'(1)}\\
M_{II} &= 1- M_{SS} - M_{SI} - \pi_S - \pi_I
\end{align}
where $\eta_{1,SS}$, $\eta_{1,SI}$, $\eta_{1,II}$, $\eta_{2,SS}$, $\eta_{2,SI}$,  and $\eta_{2,II}$ are likely to depend on $\pi_I$ and $\pi_S$ and may depend on the other quantities.  What form that dependence takes is determined by the behavior of the population.
 This completes the derivation of the equations for the actual degree formulation of serosorting.

\subsection{Variable Degree Serosorting model}
\label{sec:vd_serosort}
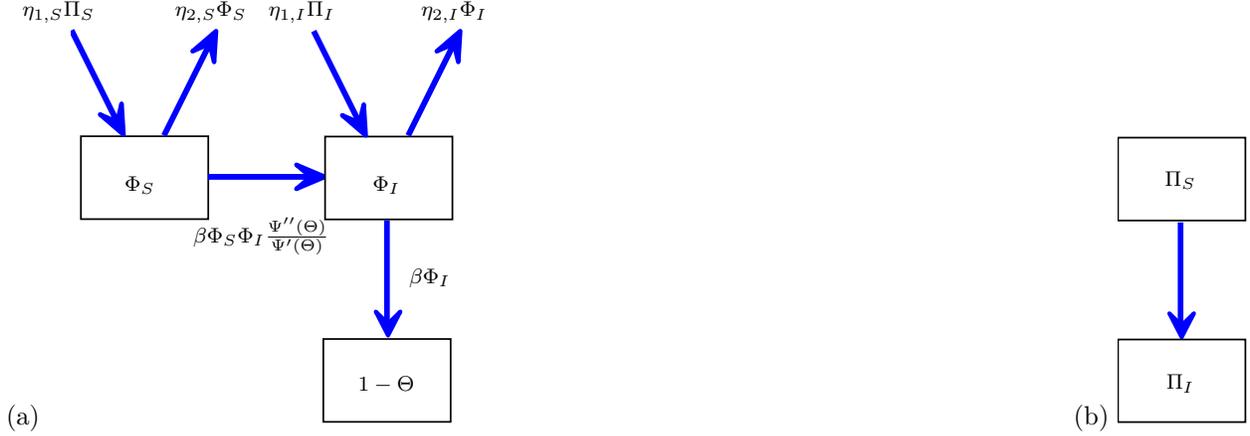
\begin{figure}
(a) \begin{picture}(0,0)%
\includegraphics{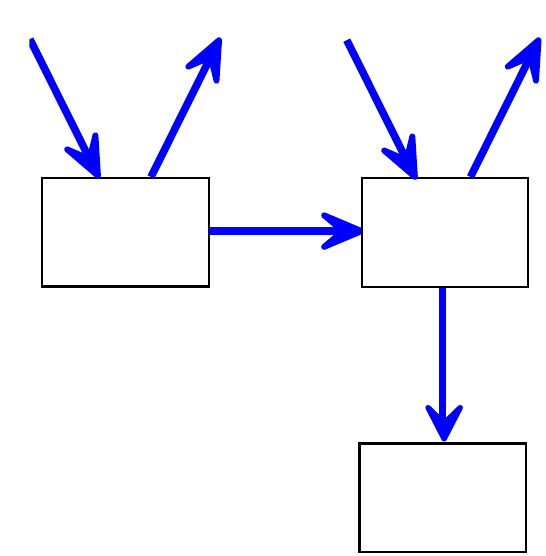}%
\end{picture}%
\setlength{\unitlength}{3947sp}%
\begingroup\makeatletter\ifx\SetFigFontNFSS\undefined%
\gdef\SetFigFontNFSS#1#2#3#4#5{%
  \reset@font\fontsize{#1}{#2pt}%
  \fontfamily{#3}\fontseries{#4}\fontshape{#5}%
  \selectfont}%
\fi\endgroup%
\begin{picture}(2608,2657)(1693,-2452)
\put(2265,-981){\makebox(0,0)[b]{\smash{{\SetFigFontNFSS{8}{9.6}{\familydefault}{\mddefault}{\updefault}{\color[rgb]{0,0,0}$\Phi_S$}%
}}}}
\put(3814,-981){\makebox(0,0)[b]{\smash{{\SetFigFontNFSS{8}{9.6}{\familydefault}{\mddefault}{\updefault}{\color[rgb]{0,0,0}$\Phi_I$}%
}}}}
\put(3814,-2238){\makebox(0,0)[b]{\smash{{\SetFigFontNFSS{8}{9.6}{\familydefault}{\mddefault}{\updefault}{\color[rgb]{0,0,0}$1-\Theta$}%
}}}}
\put(3028,-1306){\makebox(0,0)[b]{\smash{{\SetFigFontNFSS{8}{9.6}{\familydefault}{\mddefault}{\updefault}{\color[rgb]{0,0,0}$\beta\Phi_S\Phi_I\frac{\Psi''(\Theta)}{\Psi'(\Theta)}$}%
}}}}
\put(3958,-1562){\makebox(0,0)[lb]{\smash{{\SetFigFontNFSS{8}{9.6}{\familydefault}{\mddefault}{\updefault}{\color[rgb]{0,0,0}$\beta\Phi_I$}%
}}}}
\put(1755,115){\makebox(0,0)[b]{\smash{{\SetFigFontNFSS{8}{9.6}{\familydefault}{\mddefault}{\updefault}{\color[rgb]{0,0,0}$\eta_{1,S}\Pi_S$}%
}}}}
\put(2712,115){\makebox(0,0)[b]{\smash{{\SetFigFontNFSS{8}{9.6}{\familydefault}{\mddefault}{\updefault}{\color[rgb]{0,0,0}$\eta_{2,S}\Phi_S$}%
}}}}
\put(3282,115){\makebox(0,0)[b]{\smash{{\SetFigFontNFSS{8}{9.6}{\familydefault}{\mddefault}{\updefault}{\color[rgb]{0,0,0}$\eta_{1,I}\Pi_I$}%
}}}}
\put(4240,115){\makebox(0,0)[b]{\smash{{\SetFigFontNFSS{8}{9.6}{\familydefault}{\mddefault}{\updefault}{\color[rgb]{0,0,0}$\eta_{2,I}\Phi_I$}%
}}}}
\end{picture}%
\hfill\
 (b) \begin{picture}(0,0)%
\includegraphics{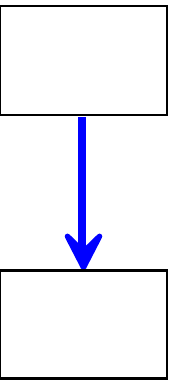}%
\end{picture}%
\setlength{\unitlength}{3947sp}%
\begingroup\makeatletter\ifx\SetFigFontNFSS\undefined%
\gdef\SetFigFontNFSS#1#2#3#4#5{%
  \reset@font\fontsize{#1}{#2pt}%
  \fontfamily{#3}\fontseries{#4}\fontshape{#5}%
  \selectfont}%
\fi\endgroup%
\begin{picture}(817,1817)(1745,-1625)
\put(2138,-123){\makebox(0,0)[b]{\smash{{\SetFigFontNFSS{8}{9.6}{\familydefault}{\mddefault}{\updefault}{\color[rgb]{0,0,0}$\Pi_S$}%
}}}}
\put(2138,-1402){\makebox(0,0)[b]{\smash{{\SetFigFontNFSS{8}{9.6}{\familydefault}{\mddefault}{\updefault}{\color[rgb]{0,0,0}$\Pi_I$}%
}}}}
\end{picture}%
\caption{\textbf{Variable-degree serosorting model.}  Flow diagram showing the interplay involved in serosorting.  We do not consider a recovered class, which simplifies the equations significantly.  The framework can be adapted to include a recovered class. The $\Pi$ variables give the proportion of contacts that would be formed with susceptible or infected individuals assuming that their behavior is not altered by disease.   The $\Phi$ variables are the probability that a current contact of the test node is with an individual of given type, under the assumption that the test node always behaves as if susceptible, and does not transmit to its neighbors.  We expect that the edge breaking and rejoining rates $\eta$ will depend on values of $\Pi_S$ and $\Pi_I$.}
\label{fig:vardeg_sero}
\end{figure}

In many populations, it is reasonable to assume that individuals create and break contacts without regard to whether contacts already exist.  Consequently, the concept of having a fixed number of stubs is inappropriate.  For these populations, we assume that in the absence of disease all contacts will have the same expected duration but different individuals will create new contacts at different rates, resulting in some having higher or lower average degrees.  In~\cite{miller:ebcm_overview}, we used $\kappa$ to be the \emph{expected degree} of a node.  However, when behavior changes based on infection status, the expected degree of individuals can change.  Instead, we refer to $\kappa$ as the \emph{desired} degree because depending on how sorting happens, it may not be possible for a node to have expected degree $\kappa$.  However, $\kappa$ will represent the expected degree of an individual if there were no infection present.  We again use $S$ and $I$ to be proportions of the population. $\Pi_S$ and $\Pi_I$ measure the proportion of desired contacts which belong to susceptible or infected individuals: $\Pi_S = \int s_\kappa \kappa \rho(\kappa) \, \mathrm{d}\kappa$, \ $\Pi_I = \int i_\kappa \kappa \rho(\kappa) \, \mathrm{d}\kappa$ where $s_\kappa$ and $i_\kappa$ denote the proportion of individuals with desired degree $\kappa$ who are susceptible and infected respectively.

We assume that the population behavior proceeds as before, but an uninfected node will end and form contacts with different rates for infected or susceptible neighbors.  There are many ways in which this could be modeled.  We will assume that a susceptible individual with desired degree $\kappa$ acquires new susceptible contacts at rate $\kappa\eta_{1,S}\Pi_S$ and new infected contacts at rate $\kappa\eta_{1,I}\Pi_I$, where both $\eta$ parameters may depend on $\Pi_S$ and $\Pi_I$.  Similarly, a susceptible individual will end an existing  contact with another susceptible and  with an infected individual at rates $\eta_{2,S}$ and $\eta_{2,I}$ respectively (where again both $\eta$ parameters may depend on $\Pi_S$ and $\Pi_I$).  We assume that $\eta_{1,S}$ and $\eta_{2,S}$ are equal if $\Pi_I=0$ so that in a disease-free population an individual's expected and desired degrees coincide.

We need to add variables in order to track the probability of having existing edges connecting to susceptible or infected nodes.  Consider a test node $u_1$ with desired degree $\kappa$, and another $u_2$ with desired degree $\kappa+\Delta \kappa$.  We define $\Phi_S\Delta\kappa$ to be the expected additional number of edges to susceptible neighbors that $u_2$ would have and $\Phi_I\Delta \kappa$ to be the expected additional number of edges to infected neighbors which have not transmitted that $u_2$ would have.  We take the values of $\Phi_S$ and $\Phi_I$ in the $\Delta \kappa \to 0$ limit.  In the cases considered in~\cite{miller:ebcm_overview}, the value of $\Phi_S$ and $\Pi_S$ were the same.  However, because there is active selection of neighbor based on disease status, in this case $\Phi_S \neq \Pi_S$.

The resulting flow diagram is shown in figure~\ref{fig:vardeg_sero}.  We must find the flux from $\Phi_S$ to $\Phi_I$.  Consider a random test node $u$ and look at a randomly chosen susceptible neighbor $v$.  Given the desired degree $\kappa_v$ of $v$, the rate that $v$ becomes infected is $\beta \Phi_I \kappa_v$.  
We need to determine the expected value of $\kappa_v$ given that $v$ is a susceptible neighbor of $u$.  We first note that the probability density function for the neighbor to be susceptible and have degree $\kappa$ is proportional to $q(\kappa)= e^{-\kappa(1-\Theta)}\kappa \rho(\kappa)/\ave{K}$ with some proportionality constant $a$.  So in order to calculate the expected value of the desired degree we take $\int a\kappa q(\kappa) \mathrm{d}\kappa/\int aq(\kappa) \mathrm{d}\kappa$.  This simplifies to $\Psi''(\Theta)/\Psi'(\Theta)$.  So the flux from $\Phi_S$ to $\Phi_I$ due to infection of the neighbor is $\beta\Phi_S\Phi_I \Psi''(\Theta)/\Psi'(\Theta)$ 
 
We find
\begin{align}
\dot{\Theta} &= -\beta \Phi_I \\
\dot{\Phi}_S &= \eta_{1,S}(\Pi_S,\Pi_I) \Pi_S - \eta_{2,S}(\Pi_S,\Pi_I) \Phi_S - \beta \Phi_S\Phi_I \frac{\Psi''(\Theta)}{\Psi'(\Theta)}\\
\dot{\Phi}_I &=  \eta_{1,I}(\Pi_S,\Pi_I) \Pi_I - [\eta_{2,I}(\Pi_S,\Pi_I) + \beta] \Phi_I + \beta \Phi_S\Phi_I \frac{\Psi''(\Theta)}{\Psi'(\Theta)}\\
\Pi_S &= \frac{\Psi'(\Theta)}{\Psi'(1)}\qquad \qquad \Pi_I = 1-\Pi_S\\
S &= \Psi(\Theta) \qquad \qquad I = 1-S
\end{align}
So an SI epidemic in a population with serosorting can be captured by a system with just three ODEs.

\section{Discussion}

We have applied the edge-based compartmental model approach introduced in~\cite{miller:ebcm_overview} to diseases and populations with different structures.  With the exception of serosorting we focused our attention on static CM networks.  We have considered each variation in isolation.  However it is possible to adapt the approach to a disease for which several of these issues are considered simultaneously in any of the network classes discussed in~\cite{miller:ebcm_overview}.

In general, we can adapt most existing mean-field/mass-action style SIR models in a closed population to the spread of infectious disease through a network.  When we do this, we get a $\phi$ variable corresponding to each of the $S$, $I$, or $R$ variables in the usual model.  We take the usual flow diagram for $S$, $I$, and $R$ and adapt it to give the fluxes between the $\phi$ variables.  We add one more compartment $1-\theta$, and flux goes from each of the potentially infectious $\phi$ variables to $1-\theta$.  This approach produces an accurate model for disease spread through the modeled population.

\section*{Acknowledgments}
JCM was supported by 1) the RAPIDD program of the Science and Technology Directorate, Department of Homeland Security and the Fogarty International Center, National Institutes of Health and 2) the Center for Communicable Disease Dynamics, Department of Epidemiology, Harvard School of Public Health under Award Number U54GM088558 from the National Institute Of General Medical Sciences.  EMV was supported by NIH K01 AI091440.  The content is solely the responsibility of the authors and does not necessarily represent the official views of the National Institute Of General Medical Sciences or the National Institutes of Health.

\bibliographystyle{plain}
\bibliography{paper3}

\end{document}